\documentclass{aa}
\usepackage{graphicx}
\begin{document}
   \title{Tracing the star formation history of cluster galaxies using the H$\alpha$/UV flux ratio}

   \author{
J. Iglesias-P\'{a}ramo
\inst{1}
\and
A. Boselli
\inst{1}
\and
G. Gavazzi
\inst{2}
\and
Antonio Zaccardo
\inst{2}
          }

\offprints{jorge.iglesias@oamp.fr}

   \institute{
Laboratoire d'Astrophysique de Marseille, Traverse du Siphon - Les
Trois Lucs, 13376 Marseille, FRANCE \\
\email{jorge.iglesias@oamp.fr,alessandro.boselli@oamp.fr}
\and
Universit\`a degli Studi di Milano - Bicocca, P.zza delle scienze 3,
20126 Milano, Italy.\\
        \email{giuseppe.gavazzi@mib.infn.it,antonio.zaccardo@mib.infn.it}
            }
   \date{Received ...; accepted ...}

\abstract{
Since the H$\alpha$ and UV fluxes from galaxies are
sensitive to stellar populations of ages $<
10^{7}$ and $\approx 10^{8}$~yr respectively, their ratio $f$(H$\alpha$)/$f$(UV)
provides us with a tool to study the recent $t \leq 10^{8}$~yr
star formation history of galaxies, an exercise that we present here applied to 98
galaxies in 4 nearby clusters (Virgo, Coma, Abell~1367 and Cancer).
The observed $f$(H$\alpha$)/$f$(UV) ratio is $\sim$ a factor of two smaller than the expected one as determined from 
population synthesis models assuming a realistic delayed, exponentially declining star formation history.
We discuss various mechanisms that may have
affected the observed $f$(H$\alpha$)/$f$(UV) ratio 
and we propose that the above discrepancy arises from
either the 
absorption of Lyman continuum photons by dust within the star formation regions 
or from the occurrence of star formation episodes. 
After splitting our sample into different subsamples according to evolutionary criteria we find that
our reference sample of galaxies unaffected by the cluster 
environment show an average value of $f$(H$\alpha$)/$f$(UV) two times lower than the expected one. 
We argue that this difference must be mostly due to absorption of $\approx 45$\% of the Lyman continuum photons within star forming regions.
Galaxies with clear signs of an ongoing interaction show average values of $f$(H$\alpha$)/$f$(UV) 
slightly higher than the reference value, as expected if those objects had SFR increased
by a factor of $\simeq 4$.
The accuracy of the current UV and H$\alpha$ photometry is not yet
sufficient to clearly disentangle the effect of interactions on the $f$(H$\alpha$)/$f$(UV) ratio,
but significant observational improvements are shortly expected
to result from the GALEX mission.  
}

   \maketitle
%

\section{Introduction}


A number of environmental mechanisms able of affecting significantly 
the evolution of galaxies in rich clusters 
have been proposed in the literature: 
gas stripping by ram
pressure (Gunn \& Gott 1972; Abadi, Moore \&
Bower 1999), galaxy-galaxy harassment in close encounters (Moore et
al. 1996), tidal stirring by the cluster potential (Byrd \& Valtonen
1990; Fujita 1998). 


These mechanisms should produce morphological
disturbances, gas removal and, on long timescales, significant 
quenching of the star formation rates (SFRs)
of galaxies due to ``fuel'' 
exhaustion (see Gavazzi et al. 2002a). However
galaxy-galaxy interactions might also enhance the
star formation in gas-rich systems, both in their nuclei and disks,
as it has been observed at several wavelengths (Larson \&
Tinsley 1978; Kennicutt et al. 1987; Hummel et al. 1987; Soifer et
al. 1984), although this {\bf enhancement} might be mild (Bergvall et al. 2003). 
The dynamical interaction of galaxies with the IGM can also produce an 
enhancement in the galaxies SFR by ram pressure (Fujita \& Nagashima 1999).

A conclusive evidence for a separate evolution of galaxies in clusters
is offered by the Butcher-Oemler effect (Butcher \& Oemler 1978), i.e. 
distant ($z \sim 0.3$) clusters show a larger fraction of blue
galaxies than nearby clusters. Several follow-up
studies (Couch \& Sharples 1987; Barger et al. 1996; Couch et al. 1994; Poggianti et
al. 1999; Balogh et al. 1999) brought to today's accepted scenario
that clusters are continuously accreting galaxies from their
neighborhood, with the accretion rate increasing with look-back time.


Several observable quantities
have been proposed as reliable estimators of the SFRs of galaxies 
(Kennicutt 1998; Rosa-Gonz\'{a}lez et al. 2002): 
H$\alpha$, UV, radio continuum and Far-IR luminosities. 
Among these, we focus our analysis on the H$\alpha$ and
UV luminosities. The H$\alpha$ luminosity comes from stars more
massive than 10M$_{\odot}$ and it traces the SFR in the last $\leq
10^{7}$~yr. The UV luminosity at 2000~\AA\ comes from stars more massive than
1.5M$_{\odot}$ and it can be used as an indicator of the SFR in the
last $\approx 10^{8}$~yr, under the condition that it stayed approximately
constant during this period. The two quantities combined, in other
words the ratio $f$(H$\alpha$)/$f$(UV), should give us a ``clock'' suitable for
telling if the SFR was constant over the last $10^{8}$~yr. 

The present paper is aimed at studying the role of the cluster environment on the star formation histories of cluster galaxies by using the $f$(H$\alpha$)/$f$(UV) ratio for a sample of galaxies in four nearby
clusters: Virgo, Coma, Abell~1367 and Cancer.
This analysis relies on the
multifrequency database that we collected so far for a large sample of 
galaxies in nearby cluster and we made available to the community
through the GOLDmine WEB site (Gavazzi et al. 2003). Beside the H$\alpha$ and UV
data which are directly used for computing the two SFR indicators,
other corollary data (e.g. Near-IR, Far-IR, H{\sc i} fluxes and optical spectroscopy) are used throughout this paper. These corollary data play a fundamental role in the determination of the dust
extinction at UV wavelengths (through the FIR/UV ratio, e.g. Buat \& Xu 1996) and at H$\alpha$ (from the Balmer decrement, e.g. Lequeux et al. 1981).

The galaxy sample is presented in Section~2. 
The observed vs. expected $f$(H$\alpha$)/$f$(UV) ratio for cluster galaxies is discussed in
Section~3. 
In Section~4 we discuss the limitations and the potentiality
of the method applied in this preliminary analysis. A
brief summary of the results is presented in Section~5. 
Details about the estimate of the birthrate parameter $b$ are given in Appendix~A. 
A second appendix contains a detailed analysis of the observational
uncertainty affecting the $f$(H$\alpha$)/$f$(UV) ratio.

\section{The sample of cluster galaxies}

The sample analyzed in this work includes late-type galaxies (morphological type later than Sa) belonging to
four nearby clusters: Virgo, Coma, A1367 and Cancer. 
Among Virgo galaxies we
selected all objects in the Virgo Cluster Catalogue
(VCC, Binggeli et al. 1985 with $m_{\mbox{\scriptsize pg}}$ $\leq$ 18) and for Coma, A1367 and Cancer
all galaxies in the 
Zwicky Catalogue (CGCG, Zwicky et al. 1961-1968 with $m_{\mbox{\scriptsize pg}}$ $\leq$ 15.7).
The accuracy of the morphological classification is excellent for the 
Virgo galaxies (Binggeli et al. 1985; 1993).
Because of the larger distances, the morphology of galaxies belonging 
to the other surveyed regions suffers from an uncertainty 
of about 1.5 Hubble type bins.
We assume a distance of 17 Mpc for the members (and possible members) 
of Virgo cluster A, 22 Mpc for 
Virgo cluster B, 32 Mpc for
objects in the M and W clouds (see Gavazzi et al. 1999).
Members of the Cancer, Coma and A1367 clusters are assumed at 
distances of 62.6, 86.6 and 92 Mpc respectively.  
Isolated galaxies in the Coma supercluster are assumed 
at their redshift distance adopting $H_{0}$ = 75 km s$^{-1}$ Mpc$^{-1}$.

\subsection{The observational dataset}

The photometric and spectroscopic data necessary for 
carrying out the present analysis (taken from
the GOLDmine database: http://goldmine.mib.infn.it/; Gavazzi et
al. 2003) are the following:
\begin{enumerate}
\item H$\alpha$ fluxes, necessary to determine the present ($\leq$ 10$^7$ years),
massive SFR (Kennicutt 1998).
H$\alpha$+[N{\sc ii}] fluxes have been obtained from imaging
(Iglesias-Paramo et al. 2002; Boselli \& Gavazzi 2002; Boselli et
al. 2002; Gavazzi et al. 2002b, and references therein): they are
integrated values and, contrary to many other samples used
for similar analysis, they do not suffer from aperture biases. 
The estimated
error on the H$\alpha$+[N{\sc ii}] flux is $\sim$ 15\%.

\item UV (2000 \AA) fluxes, useful to compute the intermediate age ($\leq 3
\times 10^{8}$ years) star formation activity (Buat et al. 1987).
The UV data are taken from the FAUST (Lampton et al. 1990) and the FOCA 
(Milliard et al. 1991) experiments. For the sake of consistency with our 
previous works, we transformed UV magnitudes taken at 1650 \AA~ by 
Deharveng et al. (1994) to 2000 \AA~ assuming a constant colour
index $m_{2000} =
m_{1650} + 0.2$~mag (see Boselli et al. 2003).
These are total magnitudes, determined by integrating the UV emission down 
to the weakest detectable isophote. The estimated error on the UV magnitude 
is 0.3 mag in general, but it ranges from 0.2 mag for bright galaxies to 0.5 
mag for weak sources observed in frames with larger than average 
calibration uncertainties.

\item Far-IR (60, 100 $\mu$m) fluxes, for obtaining an accurate UV extinction
correction (Buat et al. 2002; Boselli et al. 2003).
Far-IR at 60 and 100 $\mu$m integrated flux densities from the IRAS
survey are taken mainly from the IRAS FSC (Moshir et al. 1989). 
Three galaxies are not detected at one of these two IRAS bands and an 
upper limit is estimated to the flux: VCC~1725, 
CGCG~119-053 and CGCG~097-062. In addition, no IRAS
data are available for VCC~1699; instead, ISO data were used for this galaxy. 
The conversion between the ISO and IRAS fluxes was taken from Boselli et al. (2003).
Typical uncertainties on the Far-IR data are $\sim$ 15\%. 

\item Long slit integrated spectroscopy with detected H$\alpha$ 
and H$\beta$ lines, necessary for the determination of the Balmer 
decrement.
Long slit, drift-scan mode spectra were obtained by (Gavazzi et al. 2003b) by drifting the slit over the
whole galaxy disk, as in Kennicutt (1992).
These are intermediate ($\lambda/\Delta\lambda \sim$ 1000) resolution spectra in the
range ($3600 - 7200$ \AA). The accuracy on the determination of the
line intensities is $\approx$ 10\% for H$\alpha$ and H$\beta$ and
$\approx$ 15\% for [N{\sc ii}]$\lambda\lambda$6548,6584\AA. 

\end{enumerate}

Due to these strong observational constraints the final sample is restricted to
98 galaxies. Because of the above selection criteria, and
in particular owing to the condition that galaxies must be detected in
H$\beta$, our sample might be affected by observational biases that will be discussed in a subsequent section.

Further corollary data, when available, are used to provide information on the 
evolutionary state of the sample galaxies:

\begin{enumerate}
\item In order to quantify the degree of perturbation by the cluster-IGM, we
use the H{\sc i} deficiency parameter, defined as the logarithm of the
ratio of the observed H{\sc i} mass to the average H{\sc i} mass of isolated
objects of similar morphological type and linear size (Haynes \&
Giovanelli 1984). Galaxies with $def(\mbox{H{\sc i}}) < 0.4$ are
considered as unperturbed objects.

\item The asymmetry of the H{\sc i} profile of the individual galaxies was
also included in our analysis as an indicator of interactions in
the last $\simeq 10^{8}$~yr, as done in Gavazzi (1989). 
A galaxy with a line of sight inclination $>$ 30$^\circ$ is considered asymmetric in H{\sc i} if
its profile deviates significantly from the expected two-horns profile typical of 
unperturbed inclined galaxies, i.e. if the peak of the lowest horn is smaller than 
50 \% that of the highest one. By definition, this indicator is meaningless 
for face-on galaxies where the profile is a single horn.
H{\sc i} profiles were taken from Giovanelli \& Haynes (1985), Bothun et al. (1985), Chincarini et al. (1983), 
Helou et al. (1984), Gavazzi (1989), Haynes \& Giovanelli (1986), Hoffman et al. (1989),
Schneider et al. (1990) and Bravo-Alfaro et al. (2001).


\item Near-IR total $H$-band magnitudes are derived consistently 
with Gavazzi \& Boselli (1996) for most of the galaxies, with an accuracy of 10\%. 
For galaxies VCC~318, 459, 664, 971, 1189, 1575, 1678, 1699 and 1929, with 
no $H$-band magnitude available, it was derived from the $K$-band magnitude adopting 
$(H - K) = 0.25$ on average. The $H$-band magnitude for VCC~552 and 1091 was taken 
from the 2MASS All-Sky Extended Source Catalog (XSC). 

$H$-band luminosities are required, together with the H$\alpha$ ones, 
to estimate the birthrate parameter, $b$, defined as (Kennicutt
et al. 1994):
\begin{equation}
b = \frac{\mbox{SFR}(t)}{\left< \mbox{SFR}(t') \right>}
\end{equation}
where SFR($t$) is the SFR at the present epoch and $\left< \mbox{SFR}(t')
\right>$ is the average SFR over the galaxy lifetime. 
If we model the SF history of normal galaxies with a delayed exponential law, called ``a la Sandage''
(Gavazzi et al. 2002a), 
a value of the birthrate parameter $b_{\mbox{\scriptsize model}}$ can be estimated. On the other hand, 
an observational value of the
birthrate parameter $b_{\mbox{\scriptsize obs}}$ can be obtained from the H$\alpha$ and
$H$-band luminosities (see Boselli et al. 2001 and Appendix~A for 
details about the calculation of $b_{\mbox{\scriptsize model}}$ and $b_{\mbox{\scriptsize obs}}$). As the cluster 
environment can alter the galaxies' SFH, the ratio $b_{\mbox{\scriptsize obs}}/b_{\mbox{\scriptsize model}}$ should reflect the deviation of
the real SFH from the analytical one, and thus it should provide us with an estimate of
the effect of the environment on cluster galaxies. 
\end{enumerate}

The basic quantities used in this analysis are listed in
Table~\ref{tabla}, arranged as follows: \\
Col. (1): Galaxy name. \\
Col. (2): $\log$ of the H$\alpha$ flux corrected for dust extinction and
[N{\sc ii}] contamination as described in Section~2.2.1., in erg~s$^{-1}$~cm$^{-2}$. \\
Col. (3): $\log$ of the UV flux corrected for extinction as described
in Section~2.2.2., in erg~s$^{-1}$~cm$^{-2}$~\AA$^{-1}$. \\
Col. (4): $\log$ of the $H$-band luminosity. \\
Col. (5): H{\sc i} deficiency parameter. \\
Col. (6): Asymmetry of the H{\sc i} profile: ``S'' = symmetric,
``A'' = asymmetric and ``?''=  unclassified H{\sc i} profile.

\subsection{Extinction Correction}

\subsubsection{H$\alpha$$+$[N{\sc ii}] fluxes}

H$\alpha$$+$[N{\sc ii}] fluxes have been corrected for dust extinction
and [N{\sc ii}] contamination as in Buat et al. (2002). The integrated
spectra, available for all galaxies, have been used to
estimate the H$\alpha$/[N{\sc ii}] line ratio and the Balmer
decrement. On all spectra we were able to measure the
underlying Balmer absorption at H$\beta$. This measurement is absolutely
necessary for an accurate determination of the Balmer decrement in
intermediate star forming galaxies, where the underlying absorption
is comparable to the emission line. The average H$\beta$ equivalent width in 
absorption in our sample is 4.75~\AA.

Despite the fact that
the two [N{\sc ii}] forbidden lines are close to H$\alpha$, the triplet was 
successfully deblended in most cases by fitting three gaussian components to the ensemble
of the three lines or taking advantage of the fact that the ratio [N{\sc
ii}]$\lambda$6548/$\lambda$6584 is approximately constant.
For those galaxies for which the [N{\sc ii}]$\lambda$6548~\AA\ emission line was not detected, 
we used the theoretical relationship [N{\sc ii}]$\lambda$6548$+$$\lambda$6584~\AA\ $=$ 1.33 
$\times$ [N{\sc ii}]$\lambda$6584~\AA\ (Osterbrock 1989).

Given the
proximity of the [N{\sc ii}] doublet, deblending the underlying Balmer
absorption at H$\alpha$ results impossible. Since on  
average the equivalent width in absorption at H$\alpha$ is expected 
similar to within few percent to that of the H$\beta$ (see
Charlot \& Longhetti 2001 and references therein) its inclusion
should result in a negligible correction to the relatively strong
H$\alpha$ line.
Therefore no
correction for underlying absorption at H$\alpha$ was 
applied.

\subsubsection{UV fluxes}

UV fluxes have been corrected for galactic extinction according to
Burstein \& Heiles (1982) and for internal extinction assuming the
recipe of Boselli et al. (2003), based on the Far-IR to UV flux
ratio. This correction is, at present, the most accurate and less
model dependent, being mostly independent on the geometry, on the
SFH of galaxies and on the assumed extinction law. 

For the three galaxies with available fluxes at only one of the IRAS bands 
(60 or 100$\mu$m), the flux in the undetected IRAS band was estimated using the
templates SED of galaxies of similar luminosity given in Boselli et al. (2003).

\section{The H$\alpha$/UV ratio of star forming galaxies}

Gavazzi et al. (2002a) showed that the time evolution of optically
selected galaxies of the Virgo cluster can be
reproduced assuming an universal IMF (Salpeter) and a SFH ``a la
Sandage''. This form represents a ``delayed exponential'' SFH whose analytical 
representation as a function of time $t$ (where $t$ is the age of the galaxy) is:
\begin{equation}
\label{sandage}
\mbox{SFR}(t,\tau) \propto (t/\tau)^{2} e^{-t^{2}/2 \tau^{2}}
\end{equation}
As described in Gavazzi et al. (2002a; see their fig. 5), the temporal evolution of this
family of functions is a delayed rise of the SFR up to a maximum (at $t=\sqrt{2} \tau$), 
followed by an exponential decrease. Both the delay time and the steepness of the 
decay are regulated by a single parameter $\tau$.

The parameter $\tau$ was found to 
scale with the $H$-band luminosity, or in other words that the SFR of 
galaxies at a given time is determined by its $H$-band luminosity.
The values of $\tau$ found for our sample galaxies range from 
$3.5 \leq \tau \leq 8.5$~Gyr for normal spirals and $\tau \geq 8.5$ for
star forming dwarf galaxies of types Im and BCDs. 

For any galaxy whose spectral energy distribution (SED) is known,
the knowledge of SFR($t$) allows to predict 
the expected value (at any time $t$) of any observable quantity $A$ once we know its time evolution
by integrating over the lifetime of the galaxy:
\begin{equation}
A_{\mbox{\scriptsize exp}}(t) = \int_{0}^{t} \mbox{SFR}(t',\tau) A(t' - t) dt'
\end{equation}
where $A_{\mbox{\scriptsize exp}}(t)$ is the expected value of the variable $A$ at time $t$ and $t=0$ corresponds to the epoch of galaxy formation.

\begin{figure}[t]
   \centering
 \includegraphics[width=17cm]{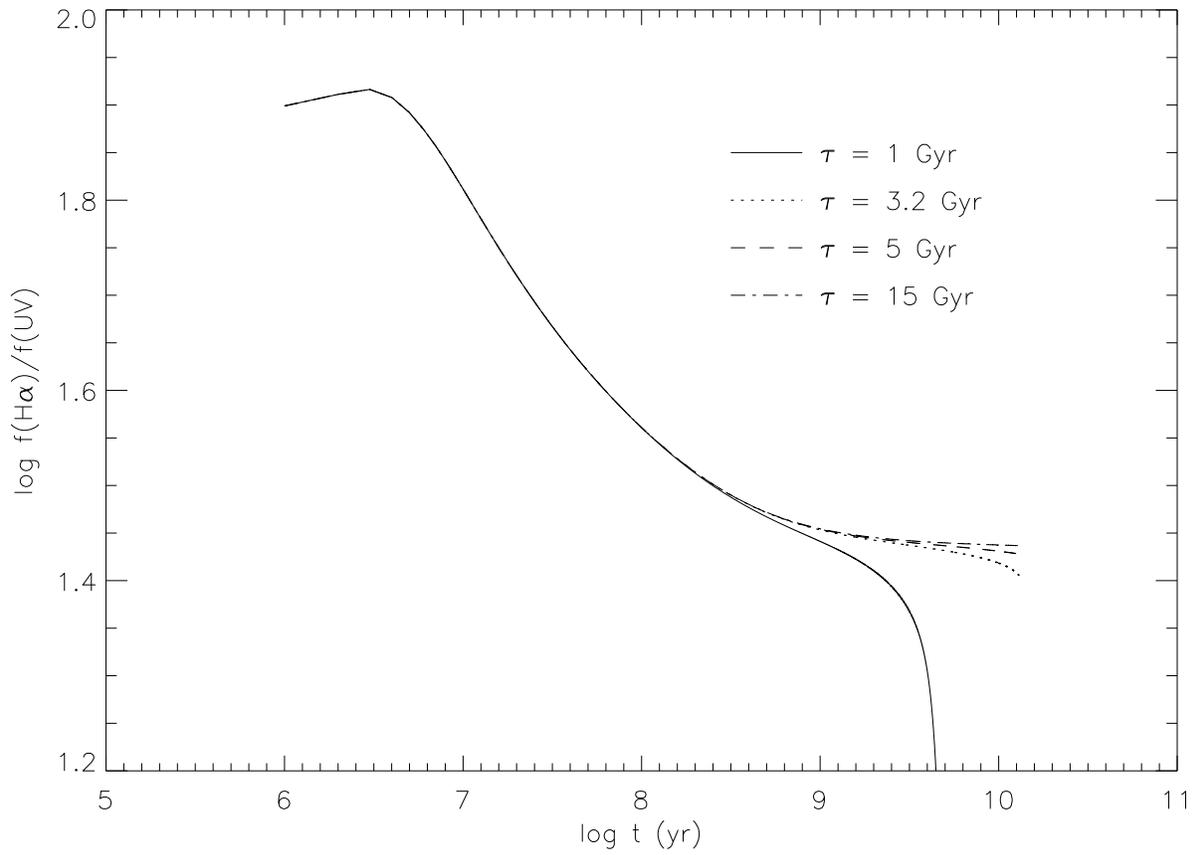}
      \caption{
Evolution of the $f$(H$\alpha$)/$f$(UV) ratio with time for galaxies
with SFH ``a la Sandage'', for values of $\tau = 1,
3.2, 5$ and $15$~Gyr. The Salpeter IMF and solar metallicity are
assumed. 
}
         \label{ha_uv_t}
   \end{figure}

Assuming eq.~\ref{sandage} for the SFH,
we show in fig.~\ref{ha_uv_t} the time evolution of the
$f$(H$\alpha$)/$f$(UV) ratio for different values of
$\tau$. The Starburst99 (Leitherer et al. 1999) code was
used, assuming a Salpeter IMF and solar
metallicity. 
As the plot shows, the ratio
$f$(H$\alpha$)/$f$(UV) shows a steep decrease in the first 1~Gyr of evolution for any $\tau$.
Between 1 and 13~Gyr the ratio
$f$(H$\alpha$)/$f$(UV) continues to decrease for $\tau = 1$~Gyr which
is typical of the brightest elliptical galaxies (see Gavazzi et
al. 2002a). 
For $\tau \geq 3$~Gyr, appropriate for normal spirals and star forming
dwarfs such as those analyzed in our work,
the ratio $f$(H$\alpha$)/$f$(UV) remains almost constant for $t \geq 10^{9}$~yr. 
Thus, if spiral galaxies follow a time evolution 
``a la Sandage'' as in eq.~\ref{sandage} 
(i.e., an almost constant SFR over the last $\approx 10^{8}$~yr), 
we expect $\log$ $f$(H$\alpha$)/$f$(UV) $\approx 1.43$ at the present time, 
according to the Starburst99 
code and assuming solar metallicity and a Salpeter IMF 
between $0.1$ and 100M$_{\odot}$.
A good agreement with this value is found when comparing the stability of this result 
to previous values reported in the literature assuming realistic SFHs and similar 
IMFs and metallicities -- 1.37 (Kennicutt et al. 1998), 1.42 (Madau et
al. (1998) and 1.51 (Boselli et al. 2001) -- so, 
it will be used as the reference value in the subsequent analysis.

\begin{figure}[t]
   \centering
 \includegraphics[width=17cm]{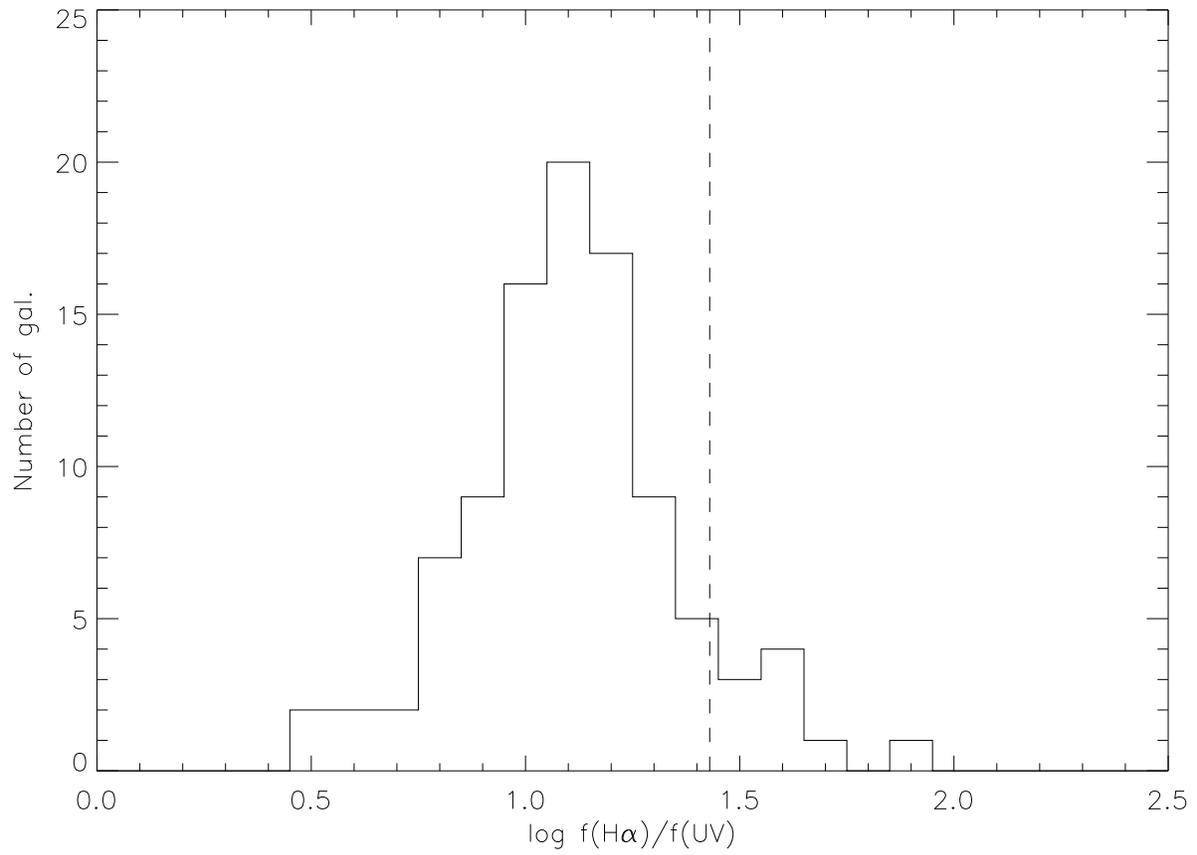}
      \caption{
Histogram of the observed $f$(H$\alpha$)/$f$(UV) ratio for galaxies in our sample. 
The dashed line corresponds to the average expected value for
evolutionary models ``a la Sandage''.
}
         \label{histo_ha_uv}
   \end{figure}

The histogram of the observed
$f$(H$\alpha$)/$f$(UV) ratio for our sample galaxies in
fig.~\ref{histo_ha_uv} shows an almost symmetric distribution centered at $\log 
f(\mbox{H}\alpha)/f(\mbox{UV})  = 1.17$ ($\sigma = 0.25$~dex), 
significantly lower than the expected value for a SFH ``a la
Sandage'' ($\log  f(\mbox{H}\alpha)/f(\mbox{UV}) = 1.43$, indicated by
the dashed line in the plot). 

The dispersion of the $f$(H$\alpha$)/$f$(UV) distribution is
consistent with that expected from the observational uncertainties, as
shown in Appendix~B. The systematic difference between  the average observed
value and the model prediction ($0.27$~dex) can hardly be explained by
systematic errors in the calibration of the data and of the
models. The nature of this difference, which we believe real,
is discussed in what follows.

\subsection{Variable IMF}

The $f$(H$\alpha$)/$f$(UV) ratio 
depends on the assumed IMF. Changing
the slope and $M_{\mbox{\scriptsize up}}$ of the IMF results in changes in the relative
numbers of the high-to-low mass stars as summarized in Table~\ref{hauvmodels}.
This table shows, for instance, the
dependence of the $f$(H$\alpha$)/$f$(UV) ratio on the IMF, assuming
instantaneous bursts and constant star formation during 
$10^{6}$, $10^{7}$ and $10^{8}$~yr (Starburst99 models). 
Three IMFs
were chosen: Salpeter ($\alpha = -2.35$ and $M_{\mbox{\scriptsize up}} = 100$M$_{\odot}$),
truncated Salpeter ($\alpha = -2.35$ and $M_{\mbox{\scriptsize up}} = 30$M$_{\odot}$), and
Miller-Scalo ($\alpha = -3.30$ and $M_{\mbox{\scriptsize up}} = 100$M$_{\odot}$; Miller \& Scalo 1979).
Not unexpectedly the Salpeter IMF gives the highest $f$(H$\alpha$)/$f$(UV)
ratio, since it corresponds to the highest high-to-low mass stars fraction.
Changing the IMF produces changes of $f$(H$\alpha$)/$f$(UV) 
of the order of $\pm 0.25$~dex for the constant SFR case. These results are quite stable
against the use of different population synthesis models: using similar initial conditions,
Starburst99 yields values of $f$(H$\alpha$)/$f$(UV) $\approx 0.07$~dex larger than PEGASE2.

\begin{figure}[t]
   \centering
 \includegraphics[width=17cm]{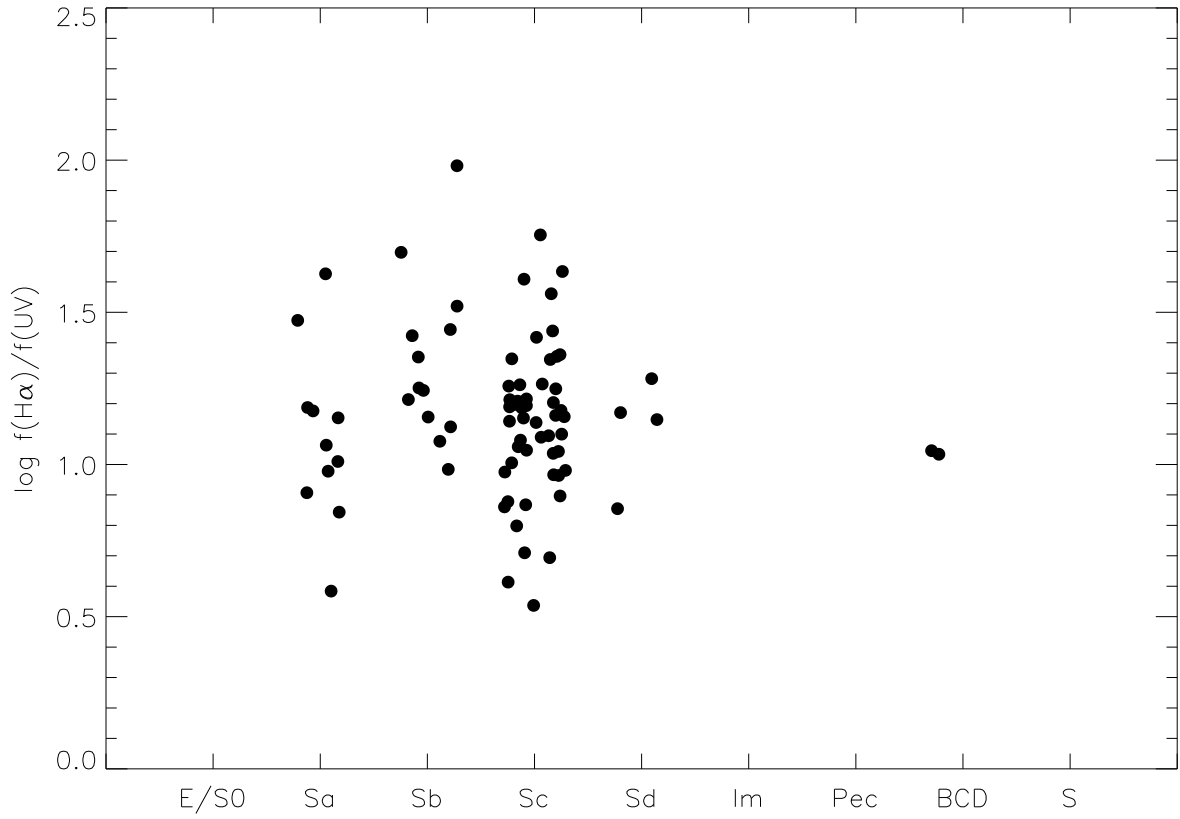}
      \caption{
$f$(H$\alpha$)/$f$(UV) ratio vs. the Hubble type for all galaxies
in our sample. A random value between $-0.4$ and 0.4 has been added to
each numerical type to avoid overplotting.
}
         \label{ha_uv_type}
   \end{figure}

From the observational point of view there is
no compelling evidence for a non universal IMF
in galaxies. The lack of any relationship between the
$f$(H$\alpha$)/$f$(UV) ratio and the morphological type or luminosity, 
as shown in fig.~\ref{ha_uv_type}, justifies the use of the same IMF for
all classes of galaxies. Moreover several studies indicate
that the Salpeter IMF for $M_{*} \geq 3$M$_{\odot}$ is adequate
for several nearby galaxies and for the Galaxy (Sakhibov \& Smirnov 2000; Massey 1998; Massey \&
Hunter 1998; see however Figer et al. 1999; Eisenhauer
et al. 1998). 

\subsection{Variable metallicity}

The $f$(H$\alpha$)/$f$(UV) ratio also depends on the metallicity of
galaxies, as shown in Table~\ref{hauvmodels}. 
The dispersion due to metallicity is maximum for an
instantaneous burst and it decreases in the case of constant star formation
over the last $10^{8}$~yr. 
The contribution of metallicity to the dispersion of the
$f$(H$\alpha$)/$f$(UV) distribution should be however minor since the
metallicities of our sample galaxies range from Z$_{\odot}$ to $0.1$Z$_{\odot}$ (see Gavazzi
et al. 2002a). 
We expect the dispersion of the $f$(H$\alpha$)/$f$(UV) distribution due
to metallicity to be $\pm 0.04$~dex around the mean theoretical
value of $\log f(\mbox{H}\alpha)/f(\mbox{UV})$, this result being independent 
on the adopted populations synthesis model. 
Thus, the systematic difference between the observed $f$(H$\alpha$)/$f$(UV)
distribution and the theoretical value can hardly be ascribed to the
different metallicities of the sample galaxies.

\subsection{Escaping of Lyman continuum photons}

A non negligible fraction of the Lyman continuum photons can escape
galaxies without ionizing hydrogen atoms. This effect would produce an overall 
shift of the $f$(H$\alpha$)/$f$(UV) distribution towards lower values. 
Indirect estimates of the escape fraction of Lyman continuum photons from H{\sc ii} regions  
determined from the ionization of the diffuse gas  by 
Zurita et al. (2000) led to values of $\sim$ 50\% in spiral discs. 
{\bf However, as pointed out by these authors, 
this has to be taken as an upper limit to the photons which escape from the galaxy since many of these ionizing photons will be absorbed by the diffuse interstellar medium so they will not escape.}
Using a similar technique, 
Bland-Hawthorn \& Maloney (1999) estimated, from the H$\alpha$ emission of the Magellanic stream,
that the escape fraction of the Milky Way is $\approx 6$\%. More direct measurements (i.e. based on
the observation of the Lyman continuum photons and not on the effect of the ionization),  
less dependent on geometrical effects, have shown that the escape of Lyman continuum photons 
from nearby starburst galaxies into the intergalactic medium is probably less than 
$\approx 10$\% (e.g. Leitherer et al. 1995, Heckman et al. 2001, Deharveng et al. 2001). 
This effect is expected to be even less important in normal galaxies than in starbursts, thus
it can be discarded as the main responsible for the
$f$(H$\alpha$)/$f$(UV) discrepancy.


\subsection{Absorption of Lyman continuum photons by dust}

Models of galaxy evolution usually assume that all Lyman continuum photons
produce the ionization of one hydrogen atom, contributing to 
the H$\alpha$ flux. However, if dust is
mixed with gas in the star formation regions, only a
fraction $f'$ of the Lyman continuum photons will encounter an hydrogen
atom, the remaining $(1 - f')$ being absorbed by the dust grains mixed with
the ionized gas. This effect, proposed by Inoue et al. (2000) 
should be properly taken into account to evaluate the energy budget of the star
formation regions, thus to calibrate the SFRs of galaxies from Far-IR fluxes.
Moreover it also produces a significant shift of the
observed $f$(H$\alpha$)/$f$(UV) ratio with respect to the model predictions.
It has been shown by Hirashita et al. (2001) that the
absorption of UV photons by dust should not depend much on
metallicity, so we can safely assume that this effect will affect in a
similar manner all galaxies in our sample. 
An average value of $f' \approx 0.57$ was found by Hirashita et
al. (2003) for a sample of galaxies similar to ours, assuming
approximately constant SFRs over the last $\approx
10^{8}$~yr. When applying this result to our sample of galaxies we obtain an almost perfect agreement between the observed and expected values of the $f$(H$\alpha$)/$f$(UV) ratio.

\subsection{Non constant SFRs \label{nonconsta}}

Galaxies with normal (i.e. ``a la Sandage'')
SFH, for any $\tau$ (fig.~1) can be assumed to have ``constant'' 
star formation over the last $10^{8}$~yr. In these conditions
the expected $f$(H$\alpha$)/$f$(UV)
is almost constant. However, Table~\ref{hauvmodels}
shows that for fixed IMFs or metallicities, non
negligible differences of $f$(H$\alpha$)/$f$(UV) are found for
different SFHs. 
It is then worthwhile to 
evaluate the consequences of 
a non-constant SFH on the $f$(H$\alpha$)/$f$(UV) ratio, which was shown in Table~\ref{hauvmodels} to produce variations on this quantity. A non-constant SFH cannot be discarded if 
bursts of star formation occurred along the evolution of galaxies.
Such events are very likely to have taken place in clusters of galaxies 
because of galaxy-galaxy and galaxy-IGM interactions.

\begin{figure}[t]
   \centering
 \includegraphics[width=17cm]{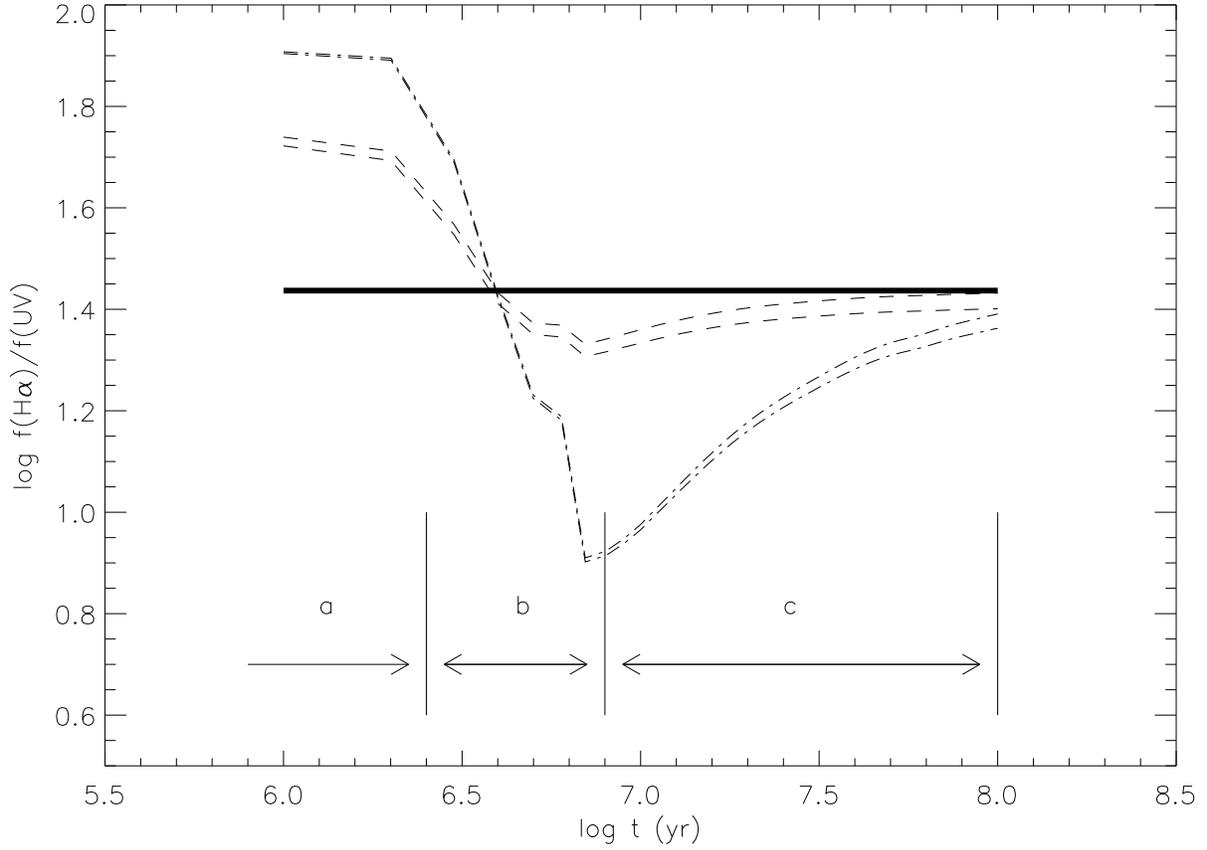}
      \caption{
Effect of instantaneous bursts of star formation on the $f$(H$\alpha$)/$f$(UV) ratio
over a normal  evolution ``a la Sandage''. The thick continuous line
represents unperturbed evolution ``a la Sandage'' for $3.2 \leq
\tau \leq 15$~Gyr (the thickness of the line accounts for the
dispersion of the models). The Salpeter IMF and solar metallicity are
assumed. 
The X axis gives the age of the instantaneous burst, assuming galaxies 13~Gyr old.
The dashed (dot dashed) lines correspond to star formation bursts of
intensities 10 (100) times the corresponding ``a la Sandage'' SFR at $t
= 13$~Gyr for $\tau = 3.2$ and $15$~Gyr.
}
         \label{mode}
   \end{figure}

Fig.~\ref{mode} shows the effect on the $f$(H$\alpha$)/$f$(UV) ratio
of instantaneous bursts of star formation superposed to the normal
evolution assumed ``a la Sandage'', with different
values of $\tau$. We have represented bursts of intensity 10 and 100
times the expected SFRs for each value of $\tau$.
One important point is that the changes in the $f$(H$\alpha$)/$f$(UV)
ratio are insensitive to $\tau$. 
The plot shows a significant increase of $f$(H$\alpha$)/$f$(UV) due to
the production of stars with $M \geq 8$M$_{\odot}$ responsible of the
H$\alpha$ emission in the first $3 \times 10^{6}$~yr (region {\bf a}), 
followed by a steep decrease as the burst fades away (region {\bf b}). Some
10$^{8}$~yr later, $f$(H$\alpha$)/$f$(UV) recovers its value previous
to the burst (region {\bf c}). The amplitude of both the
increase and the decrease of $f$(H$\alpha$)/$f$(UV) is larger for
stronger bursts. 
We remark that values of $f$(H$\alpha$)/$f$(UV)
significantly lower than the one predicted by models for constant SFR
(end of region {\bf b} in the plot) 
are reached only in the case of bursts of intensities $\geq 10$ times
the normal current SFRs of galaxies.
The presence of the strong burst of star formation
can thus account for both a shift and an
increase of the dispersion of the $f$(H$\alpha$)/$f$(UV)
distribution. 

\begin{figure}[t]
   \centering
 \includegraphics[width=17cm]{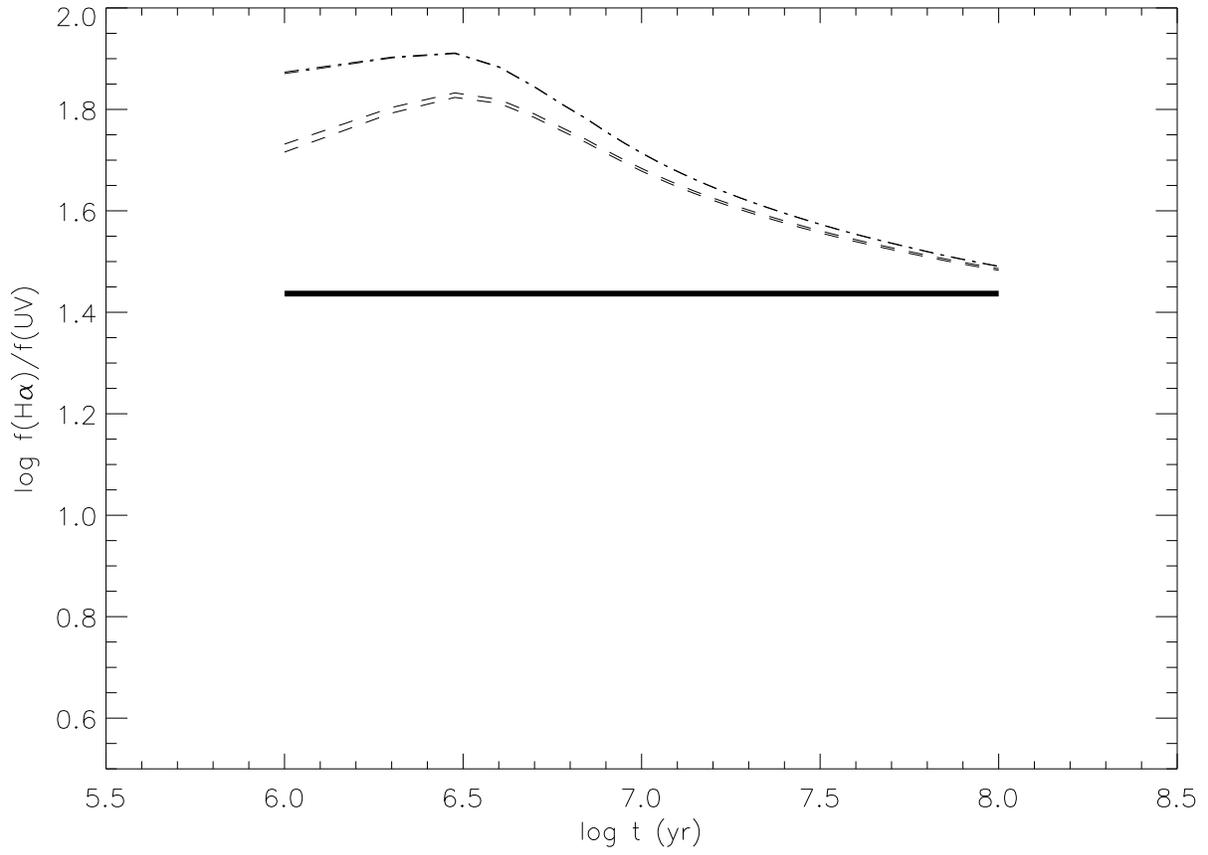}
      \caption{
Same as fig.~\ref{mode} for a burst of $10^{8}$~yr duration.
}
         \label{mode3}
   \end{figure}

The temporal dependence of the star formation
induced by galaxy interactions
is far more complex that just a single instantaneous burst 
(see Noguchi 1991 and Mihos et al. 1991,1992). 
The period over which the star formation is enhanced can last for about $10^{8}$~yr.
To show the influence of a more complex pattern of star formation on the $f$(H$\alpha$)/$f$(UV) ratio
we show in fig.~\ref{mode3} the
$f$(H$\alpha$)/$f$(UV) evolution for a burst of $10^{8}$~yr duration, overimposed to a normal evolution SFH. 
In this case, after the first $10^{7}$~yr, the $f$(H$\alpha$)/$f$(UV) ratio
decreases slowly with time and, by $10^{8}$~yr, it converges to the value for normal galaxies.

Once we know the effect of a single star formation burst on the $f$(H$\alpha$)/$f$(UV) 
ratio of a single galaxy, we simulate the expected distributions of $f$(H$\alpha$)/$f$(UV)
for a population of galaxies following a SFH
``a la Sandage'', with several overimposed star formation episodes
randomly distributed in time.
Three parameters are let free in each simulation: 
\begin{itemize}
\item the time 
over which all galaxies experience a burst of star
formation: $3 \times 10^{6}$, $10^{8}$ and $10^{9}$~yr, coincident with the timescales of 
three environmental mechanisms acting on cluster galaxies, 

\item the duration of the bursts: instantaneous
and $10^{8}$~yr, 

\item the maximum intensity of the burst: 10, 100 and 1000 times
the expected SFR for galaxies following an evolution ``a la Sandage'' at
$t = 13$~Gyr. 
\end{itemize}
In order to reproduce a more realistic variety of burst intensities
we also produced simulations in which the maximum
intensity of the bursts was randomly chosen between 0 and a certain
value (namely 10, 100 and 1000 times the normal SFR). 
By combining the above cases we have reproduced the values of the
$f$(H$\alpha$)/$f$(UV) ratio for 36 scenarios.  
An error budget consistent with the one of our dataset (detailed in Appendix~B) 
was included in the simulated H$\alpha$ and UV fluxes. 
For each scenario up to
100 simulations were run. The comparison of the resulting $f$(H$\alpha$)/$f$(UV)
distributions with the observed one are reported in Table~\ref{tabsimu}.

Let us first summarize the scenarios with instantaneous star formation bursts:
\begin{itemize}
\item For scenarios~1 to 6, where the star formation episodes are spread along the 
last $3 \times 10^{6}$~yr, the average $\log$$f$(H$\alpha$)/$f$(UV)  increases from the 
theoretical value 1.43, up to
1.91 times. All these scenarios produce distributions of $\log$$f$(H$\alpha$)/$f$(UV) non 
consistent with the observed one, as reflected by the unacceptably high $\chi_{\mbox{\scriptsize n}}^{2}$. 

\item Scenarios~7 to 12, which correspond to star formation episodes spread along $10^{8}$~yr, 
show average values of $f$(H$\alpha$)/$f$(UV) lower than the expected value (strongly depending on the
intensity of the star formation episodes). This result is expected since, as shown in fig.~\ref{mode}, 
after a strong star formation burst the value of $\log$$f$(H$\alpha$)/$f$(UV) is below the expected value and it takes
about $10^{8}$~yr to recover. It is remarkable that scenarios~9 and 12, 
which corresponds to star formation bursts of the order of 100 (1000) times the current SFRs, 
produce $f$(H$\alpha$)/$f$(UV) distributions consistent with the observed one.

\item Finally, scenarios~13 to 18, for which the star formation episodes are spread along $10^{9}$~yr, 
yield average values of $f$(H$\alpha$)/$f$(UV) slightly lower than the theoretical one. This means 
that the effect of an instantaneous burst of star formation is shorter than $10^{9}$~yr. 
None of these scenarios provide $f$(H$\alpha$)/$f$(UV) consistent with the observed one. 
\end{itemize}

Concerning the scenarios with bursts of $10^{8}$~yr duration:
\begin{itemize}
\item The behavior of scenarios 19 to 30 is similar to that of scenarios 1 to 6, 
that is, their average $f$(H$\alpha$)/$f$(UV) ratio is enhanced with respect to the value corresponding to normal
galaxies and they show a high average value of $\chi_{\mbox{\scriptsize n}}^{2}$.

\item Scenarios 31 to 36 behave like scenarios 7 to 12, with only one of them (namely scenario 33) 
been fairly consistent with the observed distribution.
\end{itemize}

Summarizing, we note that bursts of intensities about 100 
times the expected SFRs for normal galaxies are required to obtain a $f$(H$\alpha$)/$f$(UV)  
distribution consistent 
with the one observed in our sample galaxies, under the assumption that non constant star formation 
is the only mechanism governing the $f$(H$\alpha$)/$f$(UV) ratio.

\section{Discussion}

In the previous section we explored some physical mechanisms
that could possibly explain the inconsistency between the observed
distribution of $f$(H$\alpha$)/$f$(UV) and the theoretical
value. 
We reached the conclusion that, while the dispersion is
consistent with the observational uncertainties, the difference between the
average observed value of $f(\mbox{H}\alpha)/f(\mbox{UV})$ and
the theoretical value is real and might have physical
implications. Of all the explored possibilities 
only two seem able to reproduce the observed
$f$(H$\alpha$)/$f$(UV) distribution, namely:  non constant SFRs over
the last $10^{8}$~yr and the absorption of Lyman continuum photons by dust
within star forming regions. 

The non constant SFR hypothesis
has been used by Sullivan et al. (2000, 2001)
to explain the discrepancy between the observed and the theoretical H$\alpha$ and UV
fluxes in a sample of UV selected galaxies.
For galaxies in clusters, where interactions are likely to take place,
the ``non constant'' star formation scenario seems the realistic one. 
However, since not all cluster galaxies are affected by the environment in the same way, 
we split our sample in several subsamples in order study the behavior of the $f$(H$\alpha$)/$f$(UV)
distributions for galaxies in various evolutionary stages:

\begin{itemize}

\item Galaxies showing clear morphological disturbances are known to be experiencing recent 
interactions with close neighbors or with the IGM, and in most cases an enhancement of their SFRs is reflected
on their H$\alpha$ fluxes (timescale for production of 
Lyman continuum photons $\leq 10^{7}$~yr).
Three galaxies of our sample belong to this category: CGCG~097-073 and CGCG~097-079 
(Gavazzi et al. 2001a) and CGCG~097-087
(Gavazzi et al. 2001b). These
galaxies will be referred hereafter as the {\bf {\em ``interacting''}} subsample. 

\item Galaxies with asymmetric H{\sc i} profiles are known to have experienced interactions 
on timescales of $\simeq 5 \times 10^{8}$~yr (Gavazzi 1989), corresponding to the timescale necessary for
redistributing the neutral gas throughout the disk. In our sample, these
are the galaxies labeled ``A'' in last column of Table~\ref{tabla}. Hereafter, we will refer to 
them as the {\bf {\em ``asymmetric''}} subsample\footnote{In order to avoid confusion, 
we do not include the galaxies from the {\em ``interacting''} subsample in the 
{\em ``asymmetric''} subsample, although these three
galaxies show an asymmetric profile.}. Given that the timescale for removing the 
H{\sc i} asymmetries is usually larger than the timescale over which the effects of 
the interactions are apparent (i.e. close galaxy-galaxy interactions), the
enhancement of the SFRs for these galaxies is expected to be lower than for the {\em ``interacting''} ones. 

\item Another measure of the interaction with the environment is provided by the HI deficiency parameter.
As galaxies approach the cluster center they loose their peripheral gas
envelope due to ram-pressure stripping, preventing their subsequent 
star formation. The
timescale for this process is $\simeq 10^{9}$~yr, which approximately
corresponds to the cluster crossing time.  
We consider as {\bf {\em ``deficient''}} galaxies those with $def(\mbox{H{\sc i}}) \geq 0.4$. 
We exclude from this
subsample deficient galaxies with asymmetric profiles, in order separate the effects of H{\sc i} 
deficiency from interactions.

\item Finally, we define a {\bf {\em ``reference''}} sample of galaxies for which no traces of 
interaction with the cluster environment are found: they have a normal H{\sc i} content (i.e., $def(\mbox{H{\sc
i}}) < 0.4$) and do not show neither clear signatures of interactions nor asymmetric 
H{\sc i} profiles. These galaxies will be considered hereafter as ``normal'' galaxies.

\end{itemize}

Table~\ref{tabmedia} lists the average values of the
$f$(H$\alpha$)/$f$(UV) and $b_{\mbox{\scriptsize obs}}/b_{\mbox{\scriptsize model}}$ ratios for each 
different subsample. 
Figs.~\ref{ha_uv_hidef} and \ref{dbirth_hidef} show the $f$(H$\alpha$)/$f$(UV) and 
$b_{\mbox{\scriptsize obs}}/b_{\mbox{\scriptsize model}}$ ratios of the individual galaxies of the three subsamples vs. their H{\sc i} deficiency. 
    
The {\em ``reference''} galaxies show $\left< \log b_{\mbox{\scriptsize obs}}/b_{\mbox{\scriptsize model}} \right> = 0$, meaning 
that their recent star formation activity coincides with the expected one. 
In addition, $\left< \log f(\mbox{H}\alpha)/f(\mbox{UV}) \right> = 1.11$, which does not correspond to 
the theoretical value of 1.43 predicted by synthesis models. Given that these galaxies
are selected for their normal H{\sc i} content and no traces of interactions, we take 
this value as a reference value for normal star forming galaxies. Absorption of $\simeq 45$\% 
Lyman continuum photons by
dust within H{\sc ii} regions should  account for the discrepancy between the observed 
and theoretical value of the
$f$(H$\alpha$)/$f$(UV) ratio for normal galaxies. 

Moving on to galaxies perturbed by the cluster environment, we find that the 
{\em ``interacting''} and {\em ``asymmetric''} galaxies show values of $\log$$f$(H$\alpha$)/$f$(UV) 0.14~dex
higher than {\em ``reference''} galaxies. As we showed in Section~3.5, 
the presence of star formation bursts is likely to produce such an enhancement. 
From the $b_{\mbox{\scriptsize obs}}/b_{\mbox{\scriptsize model}}$
ratio we estimate that the intensity of the star formation activity 
is at present 3.5 times higher for the {\em
``interacting''} galaxies compared to the {\em asymmetric} and the {\em ``reference''} ones.
Given that the {\em ``interacting''} galaxies are presently undergoing an 
interaction, the age of the burst is $\approx 10^{6}$~yr, thus the increase of 
the star formation activity is maximal (see fig.~\ref{mode}). We thus expect that,
consistently with model predictions (Fujita 1998),
galaxy--galaxy or galaxy--IGM interactions in clusters can induce bursts 
of star formation able to increase by up to a factor of $\approx 4$ the expected 
SFR of normal late-type galaxies.  



Finally, we analyze the behavior of the {\em ``deficient''} galaxies. 
These galaxies have been shown to have lower than expected star formation 
activity as measured by the $b$ parameter
(Boselli et al. 2001). However, we find for them higher  
$f$(H$\alpha$)/$f$(UV) and of $b_{\mbox{\scriptsize obs}}/b_{\mbox{\scriptsize model}}$ than for the {\em ``reference''} galaxies. 
This apparent contradiction might be due to selection effects.
To illustrate this point we show in Table~\ref{bias} the average values of 
$EW(\mbox{H}\alpha + \mbox{[N{\sc ii}]})$ for subsamples of galaxies satisfying the 
various observational constraints, separately for
deficient and non deficient galaxies. It appears that, as more 
observational constraints are applied, the resulting average 
$EW(\mbox{H}\alpha + \mbox{[N{\sc ii}]})$ tends to increase, biasing towards 
more actively star forming galaxies. 
For the non deficient galaxies, this bias affects the average $EW(\mbox{H}\alpha + \mbox{[N{\sc ii}]})$ 
by less than 30\%. For the
deficient galaxies, by imposing the condition for H$\beta$ line detection,  
the estimate of $EW(\mbox{H}\alpha + \mbox{[N{\sc ii}]})$ results doubled.  
The {\em ``reference''}, 
{\em ``interacting''} and {\em ``asymmetric''} samples are less affected by this selection bias because they
contain non-deficient objects.


\section{Conclusions}

The $f$(H$\alpha$)/$f$(UV) ratio of cluster galaxies is analyzed in this paper
as a promising tool to estimate if the the star formation history of galaxies 
has remained constant on timescales of $\simeq 10^{8}$~yr.
The observed $f$(H$\alpha$)/$f$(UV) distribution is compared to the one predicted
by models of galaxies, assuming a continuum SFH. The
dispersion of the observed $f$(H$\alpha$)/$f$(UV) distribution is
consistent with the one expected from the observational uncertainties. 
We find a systematic negative difference between the average observed value and the
model predictions.
We discuss some mechanisms that could possibly produce such an
observed difference and we highlight the two most likely ones: 
the absorption and, in a minor way, the escape of Lyman continuum photons 
and the occurrence of star formation bursts overimposed to a
smooth SFH. 

The $f$(H$\alpha$)/$f$(UV)
distribution is considered for different galaxy subsamples, each
of them comprising galaxies in different evolutionary stages, 
possibly induced by the cluster environment. 
The {\em ``reference''} unperturbed galaxies have $f$(H$\alpha$)/$f$(UV) 
lower by 0.34 dex on average than the one predicted by the models. We  
suggest that absorption (and to a lesser extent escape) of Lyman continuum photons 
causes the observed discrepancy. We estimate that 
about 45\% of the Lyman continuum photons are absorbed
by dust in the star forming regions before ionization, consistently with the estimate 
of Hirashita et al. (2003) on similar objects.

When galaxies with signatures of recent or past 
interactions with the cluster environment ({\em ``interacting''} and {\em ``asymmetric''}) 
are considered, we find that their
$f$(H$\alpha$)/$f$(UV) ratio is slightly higher than the one of {\em ``reference''} galaxies.  
Even though the absorption of Lyman continuum photons is taken into account,
the observed $f$(H$\alpha$)/$f$(UV) ratio can be reconciled to the predicted one
only assuming that these objects underwent bursts of star formation 
of intensity  $\sim$ 100 times larger than normal, as intense as Arp~220.
Objects of this kind are however not presently observed in nearby clusters.

The present observational uncertainties on both the H$\alpha$ and UV fluxes
are still too large to allow disentangling the effects of recent star
formation bursts from those of absorption of Lyman continuum photons. 
However we stress the potentiality of the proposed H$\alpha$/UV method for 
studying the recent history of star formation in late type galaxies, once
improvements in modeling the radiation transfer
through the dust in star forming regions will be achieved and more precise
UV and Far-IR photometry will be available. This will soon become a reality after the GALEX and ASTRO-F
experiments will perform their all sky surveys, providing $\Delta f(\mbox{UV})$ and $\Delta
f(\mbox{Far-IR}) \approx 10$\%. 

\begin{acknowledgements}
We thank Veronique Buat and Jean Michel Deharveng for interesting comments and suggestions.
JIP acknowledges the Fifth Framework Program of the EU for a Marie
Curie Postdoctoral Fellowship.
This research has made use of the NASA/IPAC Extragalactic Database (NED) and of the NASA/ 
IPAC Infrared Science Archive, which are operated by the Jet Propulsion Laboratory, California Institute of
Technology, under contract with the National Aeronautics and Space Administration. 
We also acknowledge the GOLD Mine Database, operated by the Universit\'{a} degli 
Studi di Milano-Bicocca. 
\end{acknowledgements}

\newpage

\onecolumn

\begin{table}
      \caption[]{Basic properties and observational data of the sample galaxies.}
         \label{tabla}
     $$ 
         \begin{tabular}{lccccc}
            \hline
            \noalign{\smallskip}
Name & $\log f(\mbox{H}\alpha)$ & $\log f(\mbox{UV})$ & $\log L_{\mbox{\scriptsize H}}$ &
$def(\mbox{H{\sc i}})$ & H{\sc i} asymm. \\
     & (erg~s$^{-1}$~cm$^{-2}$) & (erg~s$^{-1}$~cm$^{-2}$~\AA$^{-1}$) & (L$_{\odot}$) & & \\
            \noalign{\smallskip}
            \hline
            \noalign{\smallskip}
VCC     25 & $-$11.59 & $-$12.75 & 10.39 & $-$0.21 & ? \\
VCC     66 & $-$11.48 & $-$12.89 & 10.22 & $-$0.20 & ? \\
VCC     89 & $-$11.74 & $-$12.78 & 10.65 & $-$0.03 & S \\
VCC     92 & $-$11.21 & $-$12.33 & 10.99 &  0.33 & ? \\
VCC    131 & $-$12.60 & $-$13.30 &  9.52 &  0.09 & S \\
VCC    157 & $-$11.31 & $-$12.65 & 10.48 &  0.61 & S \\
VCC    221 & $-$11.94 & $-$13.12 &  9.88 &  0.41 & S \\
VCC    307 & $-$10.73 & $-$11.95 & 10.94 &  0.01 & S \\
VCC    318 & $-$12.48 & $-$13.68 &  9.18 & $-$0.13 & S \\
VCC    382 & $-$12.05 & $-$12.59 & 10.65 & $-$0.32 & S \\
VCC    459 & $-$12.58 & $-$13.61 &  8.73 & $-$0.07 & S \\
VCC    491 & $-$11.75 & $-$12.89 &  9.42 & $-$0.29 & S \\
VCC    508 & $-$10.72 & $-$11.86 & 10.98 & $-$0.06 & S \\
VCC    552 & $-$12.23 & $-$13.29 &  8.99 & $-$0.42 & S \\
VCC    664 & $-$12.17 & $-$13.38 &  8.92 &  0.62 & S \\
VCC    667 & $-$12.75 & $-$13.71 &  9.78 &  0.58 & S \\
VCC    692 & $-$12.50 & $-$13.36 &  9.64 &  0.66 & S \\
VCC    699 & $-$12.37 & $-$13.31 &  9.71 &  0.19 & S \\
VCC    787 & $-$12.47 & $-$13.35 &  9.65 &  0.26 & S \\
VCC    801 & $-$11.58 & $-$13.12 &  9.97 & $-$0.62 & S \\
VCC    827 & $-$12.10 & $-$13.15 & 10.14 &  0.08 & S \\
VCC    836 & $-$11.51 & $-$12.57 & 10.54 &  0.69 & S \\
VCC    849 & $-$12.11 & $-$13.36 &  9.79 &  0.41 & S \\
VCC    851 & $-$12.18 & $-$13.79 &  9.80 &  0.23 & A \\
VCC    865 & $-$11.86 & $-$13.12 &  9.69 &  0.38 & S \\
VCC    873 & $-$11.15 & $-$12.79 & 10.39 &  0.63 & S \\
VCC    905 & $-$12.63 & $-$13.59 &  9.66 &  0.35 & S \\
VCC    912 & $-$12.19 & $-$13.26 &  9.88 &  0.99 & S \\
VCC    921 & $-$11.89 & $-$13.05 &  9.64 &  0.59 & S \\
VCC    938 & $-$12.09 & $-$13.26 &  9.74 &  0.36 & S \\
VCC    939 & $-$12.32 & $-$13.18 &  9.87 &  0.24 & S \\
VCC    957 & $-$11.64 & $-$12.90 &  9.91 &  0.02 & ? \\
VCC    971 & $-$12.40 & $-$13.55 &  9.50 &  0.20 & ? \\
VCC    979 & $-$11.98 & $-$13.14 & 10.45 &  1.17 & S \\
VCC    980 & $-$12.52 & $-$13.55 &  8.77 &  0.67 & ? \\
VCC   1002 & $-$11.65 & $-$13.21 & 10.22 &  0.47 & S \\
VCC   1091 & $-$12.22 & $-$13.47 &  8.86 & $-$0.35 & S \\
VCC   1118 & $-$12.08 & $-$13.27 & 10.08 &  0.51 & S \\
VCC   1189 & $-$12.55 & $-$13.65 &  9.25 &  0.34 & A \\
VCC   1193 & $-$12.43 & $-$13.79 &  9.28 & $-$0.05 & S \\
VCC   1205 & $-$12.41 & $-$13.02 &  9.73 & $-$0.03 & S \\
VCC   1290 & $-$12.20 & $-$13.18 &  9.91 &  0.05 & S \\
VCC   1379 & $-$12.09 & $-$13.07 &  9.84 &  0.15 & S \\
VCC   1393 & $-$12.22 & $-$13.32 &  9.47 &  0.23 & S \\
VCC   1401 & $-$10.76 & $-$12.20 & 11.18 &  0.55 & S \\
VCC   1450 & $-$12.01 & $-$12.99 &  9.47 &  0.54 & S \\
VCC   1508 & $-$11.59 & $-$12.75 &  9.93 & $-$0.26 & S \\
VCC   1516 & $-$11.77 & $-$13.19 &  9.85 &  0.80 & S \\
VCC   1532 & $-$12.35 & $-$13.44 &  9.55 &  0.82 & S \\
VCC   1554 & $-$11.28 & $-$12.56 &  9.90 & $-$0.37 & S \\
            \noalign{\smallskip}
            \hline
         \end{tabular}
     $$ 
   \end{table}

\clearpage

\addtocounter{table}{-1}

\begin{table}
      \caption[]{Continued.}
     $$ 
         \begin{tabular}{lccccc}
            \hline
            \noalign{\smallskip}
Name & $\log f(\mbox{H}\alpha)$ & $\log f(\mbox{UV})$ & $\log L_{\mbox{\scriptsize H}}$ &
$def(\mbox{H{\sc i}})$ & H{\sc i} asymm. \\
     & (erg~s$^{-1}$~cm$^{-2}$) & (erg~s$^{-1}$~cm$^{-2}$~\AA$^{-1}$) & (L$_{\odot}$) & & \\
            \noalign{\smallskip}
            \hline
            \noalign{\smallskip}
VCC   1575 & $-$11.08 & $-$12.27 & 10.76 &  0.19 & S \\
VCC   1588 & $-$12.04 & $-$13.05 & 10.05 &  0.68 & S \\
VCC   1678 & $-$12.45 & $-$13.62 &  8.81 & $-$0.06 & S \\
VCC   1699 & $-$12.90 & $-$13.75 &  8.83 &  0.04 & S \\
VCC   1725 & $-$12.70 & $-$13.74 &  8.97 &  0.55 & S \\
VCC   1811 & $-$12.31 & $-$13.21 &  9.85 &  0.23 & S \\
VCC   1929 & $-$12.63 & $-$13.34 &  9.51 &  0.35 & S \\
VCC   1943 & $-$11.43 & $-$12.95 & 10.26 &  0.25 & S \\
VCC   1972 & $-$11.32 & $-$12.66 & 10.50 &  0.27 & S \\
VCC   1987 & $-$11.23 & $-$12.31 & 10.66 & $-$0.29 & A \\
VCC   2058 & $-$11.15 & $-$12.91 & 10.35 &  0.90 & S \\
CGCG 043-034 & $-$11.42 & $-$12.86 &  9.79 & $-$0.29 & S \\
CGCG 043-071 & $-$11.32 & $-$12.68 & 10.12 & $-$0.75 & S \\
CGCG 043-093 & $-$11.43 & $-$12.64 & 10.27 & $-$0.07 & S \\
CGCG 097-062 & $-$12.97 & $-$14.30 & 10.10 &  0.31 & A \\
CGCG 097-068 & $-$11.74 & $-$13.43 & 10.78 & $-$0.14 & S \\
CGCG 097-073 & $-$12.77 & $-$14.07 & 10.00 &  0.16 & A \\
CGCG 097-079 & $-$12.70 & $-$13.87 & 10.02 &  0.25 & A \\
CGCG 097-087 & $-$11.97 & $-$13.25 & 10.88 &  0.19 & A \\
CGCG 097-091 & $-$12.69 & $-$13.67 & 10.85 & $-$0.18 & S \\
CGCG 097-120 & $-$12.13 & $-$13.75 & 11.06 &  0.90 & S \\
CGCG 100-004 & $-$11.33 & $-$12.60 & 10.52 & $-$0.24 & S \\
CGCG 119-029 & $-$12.13 & $-$13.34 & 10.75 & $-$0.30 & S \\
CGCG 119-041 & $-$12.83 & $-$13.74 & 10.51 &  0.30 & S \\
CGCG 119-043 & $-$12.70 & $-$13.94 & 10.07 &  0.29 & S \\
CGCG 119-046 & $-$12.15 & $-$13.39 & 10.32 & $-$0.22 & S \\
CGCG 119-047 & $-$12.42 & $-$13.43 & 10.42 & $-$0.61 & S \\
CGCG 119-053 & $-$13.00 & $-$13.98 & 10.12 & $-$0.37 & S \\
CGCG 119-054 & $-$12.44 & $-$13.91 & 10.70 &  ---  & ? \\
CGCG 119-059 & $-$13.13 & $-$13.93 &  9.66 &  0.14 & S \\
CGCG 119-068 & $-$12.66 & $-$13.84 & 10.40 & $-$0.27 & S \\
CGCG 119-085 & $-$13.61 & $-$14.19 & 10.38 & $-$0.17 & S \\
CGCG 127-049 & $-$12.52 & $-$13.88 & 10.51 &  0.32 & S \\
CGCG 160-020 & $-$13.00 & $-$13.82 &  9.98 &  0.27 & S \\
CGCG 160-026 & $-$12.96 & $-$14.11 & 10.31 &  0.23 & A \\
CGCG 160-055 & $-$12.56 & $-$13.41 & 10.96 &  0.49 & A \\
CGCG 160-058 & $-$12.65 & $-$14.00 & 10.58 &  0.40 & S \\
CGCG 160-067 & $-$12.62 & $-$13.86 & 10.06 & $-$0.05 & S \\
CGCG 160-076 & $-$12.88 & $-$14.08 &  9.87 & $-$0.35 & S \\
CGCG 160-086 & $-$12.84 & $-$14.09 & 10.05 &  0.76 & ? \\
CGCG 160-088 & $-$12.06 & $-$14.04 & 10.88 &  0.42 & S \\
CGCG 160-106 & $-$12.55 & $-$13.90 & 10.99 &  0.54 & ? \\
CGCG 160-108 & $-$12.92 & $-$14.09 & 10.16 &  ---  & ? \\
CGCG 160-128 & $-$12.73 & $-$13.91 &  9.86 &  ---  & ? \\
CGCG 160-139 & $-$12.60 & $-$13.81 & 10.02 & $-$0.19 & A \\
CGCG 160-213 & $-$12.72 & $-$13.81 & 10.10 &  ---  & ? \\
CGCG 160-252 & $-$12.49 & $-$13.55 & 10.38 &  0.56 & S \\
CGCG 160-260 & $-$12.47 & $-$13.64 & 11.21 &  0.81 & S \\
            \noalign{\smallskip}
            \hline
         \end{tabular}
     $$ 
   \end{table}

\clearpage

   \begin{table}
      \caption[]{Dependence of the $f$(H$\alpha$)/$f$(UV) ratio on the IMF parameters, metallicity and star formation history: (1) Evolutionary synthesis code; (2) Metallicity; (3) IMF slope; (4)
Lower limit for the IMF; (5) Upper limit for the IMF; (6) Time interval over which the SFR is considered constant; (7) $\log$ of the $f$(H$\alpha$)/$f$(UV) ratio.}
         \label{hauvmodels}
     $$ 
         \begin{tabular}{lcccccc}
            \hline
            \noalign{\smallskip}
Source & $Z$ & IMF & $M_{\mbox{\scriptsize low}}$ & $M_{\mbox{\scriptsize up}}$ & $t$ & $\log$ $f$(H$\alpha$)/$f$(UV) \\
            \noalign{\smallskip}
            \hline
            \noalign{\smallskip}
PEGASE2 & 0.0004 & Salpeter & 1 & 100 & $10^{7}$ & 1.79 \\
PEGASE2 & 0.004  & Salpeter & 1 & 100 & $10^{7}$ & 1.69 \\
PEGASE2 & 0.02   & Salpeter & 1 & 100 & $10^{7}$ & 1.51 \\
PEGASE2 & 0.05   & Salpeter & 1 & 100 & $10^{7}$ & 1.35 \\
Starburst99 & 0.001 & Salpeter & 1 & 100 & $10^{7}$ & 1.75 \\
Starburst99 & 0.004 & Salpeter & 1 & 100 & $10^{7}$ & 1.67 \\
Starburst99 & 0.008 & Salpeter & 1 & 100 & $10^{7}$ & 1.63 \\
Starburst99 & 0.020 & Salpeter & 1 & 100 & $10^{7}$ & 1.57 \\
Starburst99 & 0.040 & Salpeter & 1 & 100 & $10^{7}$ & 1.51 \\
\hline
PEGASE2 & 0.0004 & Salpeter & 1 & 100 & $10^{8}$ & 1.54 \\
PEGASE2 & 0.004  & Salpeter & 1 & 100 & $10^{8}$ & 1.46 \\
PEGASE2 & 0.02   & Salpeter & 1 & 100 & $10^{8}$ & 1.33 \\
PEGASE2 & 0.05   & Salpeter & 1 & 100 & $10^{8}$ & 1.20 \\
Starburst99 & 0.001 & Salpeter & 1 & 100 & $10^{8}$ & 1.50 \\
Starburst99 & 0.004 & Salpeter & 1 & 100 & $10^{8}$ & 1.44 \\
Starburst99 & 0.008 & Salpeter & 1 & 100 & $10^{8}$ & 1.42 \\
Starburst99 & 0.020 & Salpeter & 1 & 100 & $10^{8}$ & 1.38 \\
Starburst99 & 0.040 & Salpeter & 1 & 100 & $10^{8}$ & 1.34 \\
\hline
Starburst99 & 0.001 & Salpeter & 1 & 30 & $10^{8}$ & 1.04 \\
Starburst99 & 0.004 & Salpeter & 1 & 30 & $10^{8}$ & 0.93 \\
Starburst99 & 0.008 & Salpeter & 1 & 30 & $10^{8}$ & 0.88 \\
Starburst99 & 0.020 & Salpeter & 1 & 30 & $10^{8}$ & 0.80 \\
Starburst99 & 0.040 & Salpeter & 1 & 30 & $10^{8}$ & 0.78 \\
Starburst99 & 0.001 & Miller-Scalo & 1 & 100 & $10^{8}$ & 0.95 \\
Starburst99 & 0.004 & Miller-Scalo & 1 & 100 & $10^{8}$ & 0.89 \\
Starburst99 & 0.008 & Miller-Scalo & 1 & 100 & $10^{8}$ & 0.86 \\
Starburst99 & 0.020 & Miller-Scalo & 1 & 100 & $10^{8}$ & 0.82 \\
Starburst99 & 0.040 & Miller-Scalo & 1 & 100 & $10^{8}$ & 0.78 \\
            \noalign{\smallskip}
            \hline
         \end{tabular}
     $$ 
   \end{table}

\clearpage

   \begin{table}
      \caption[]{The simulated scenarios for 
instantaneous bursts of star formation: (1) Identificator of the model; (2) Time interval over which all simulated galaxies
experience a burst of star formation (yr); (3) Duration of the burst (yr); (4) Maximum intensity of the burst in units of the
expected SFR of galaxies following an evolution ``a la Sandage'' at $t = 13$~Gyr; (5) Average $\chi^{2}_{\mbox{\scriptsize n}}$
between the observed and each of the simulated distributions; (6) Average value of $f$(H$\alpha$)/$f$(UV) for 100
simulated distributions; (7) Average of $\sigma$$f$(H$\alpha$)/$f$(UV) for 100
simulated distributions.}
         \label{tabsimu}
     $$ 
         \begin{tabular}{rcccccc}
            \hline
            \noalign{\smallskip}
Id. & $\Delta t$ & Duration & Intensity &
$\left<\chi^{2}_{\mbox{\scriptsize n}}\right>$ & $\left<\log
f(\mbox{H}\alpha)/f(\mbox{UV})\right>$ & $\left< \sigma
\right>$ \\
            \noalign{\smallskip}
            \hline
            \noalign{\smallskip}
1 & $3 \times 10^{6}$ & Inst. & 10 for all galaxies         & $7.39\pm0.98$ & 1.71
& 0.22 \\
2 & $3 \times 10^{6}$ & Inst. & Random between 0 and 10     & $6.20\pm0.98$ & 1.58
& 0.26 \\
3 & $3 \times 10^{6}$ & Inst. & 100 for all galaxies       & $6.17\pm1.44$ & 1.90
& 0.23 \\
4 & $3 \times 10^{6}$ & Inst. & Random between 0 and 100   & $6.94\pm1.13$ & 1.81
& 0.24 \\
5 & $3 \times 10^{6}$ & Inst. & 1000 for all galaxies     & $6.04\pm1.47$ & 1.91
& 0.23 \\
6 & $3 \times 10^{6}$ & Inst. & Random between 0 and 1000 & $5.88\pm1.25$ & 1.91
& 0.26 \\
7 & $10^{8}$ & Inst. & 10 for all galaxies         & $2.62\pm0.70$ & 1.41
& 0.26 \\
8 & $10^{8}$ & Inst. & Random between 0 and 10     & $2.71\pm0.61$ & 1.39
& 0.25 \\
9 & $10^{8}$ & Inst. & 100 for all galaxies       & $0.87\pm0.38$ & 1.25
& 0.28 \\
10 & $10^{8}$ & Inst. & Random between 0 and 100   & $1.53\pm0.48$ &
1.37 & 0.26 \\
11 & $10^{8}$ & Inst. & 1000 for all galaxies     & $3.03\pm0.74$ &
0.92 & 0.42 \\
12 & $10^{8}$ & Inst. & Random between 0 and 1000 & $1.03\pm0.38$ &
1.12 & 0.39 \\
13 & $10^{9}$ & Inst. & 10 for all galaxies         & $2.67\pm0.62$ &
1.35 & 0.23 \\
14 & $10^{9}$ & Inst. & Random between 0 and 10     & $2.84\pm0.66$ &
1.41 & 0.22 \\
15 & $10^{9}$ & Inst. & 100 for all galaxies       & $2.46\pm0.66$ &
1.35 & 0.23 \\
16 & $10^{9}$ & Inst. & Random between 0 and 100   & $2.68\pm0.74$ &
1.34 & 0.23 \\
17 & $10^{9}$ & Inst. & 1000 for all galaxies     & $1.59\pm0.53$ &
1.31 & 0.34 \\
18 & $10^{9}$ & Inst. & Random between 0 and 1000 & $1.84\pm0.45$ &
1.39 & 0.27 \\
19 & $3 \times 10^{6}$ & $10^{8}$ & 10 for all galaxies & $7.53\pm0.96$ & 1.73 & 0.26 \\
20 & $3 \times 10^{6}$ & $10^{8}$ & Random between 0 and 10 & $4.88\pm0.78$ & 1.53 & 0.27 \\
21 & $3 \times 10^{6}$ & $10^{8}$ & 100 for all galaxies & $6.35\pm1.17$ & 1.88 & 0.25 \\
22 & $3 \times 10^{6}$ & $10^{8}$ & Random between 0 and 100 & $6.87\pm1.23$ & 1.82 & 0.24 \\
23 & $3 \times 10^{6}$ & $10^{8}$ & 1000 for all galaxies & $6.05\pm1.31$ & 1.91 & 0.24 \\
24 & $3 \times 10^{6}$ & $10^{8}$ & Random between 0 and 1000 & $6.04\pm1.42$ & 1.92 & 0.24 \\
25 & $10^{8}$ & $10^{8}$ & 10 for all galaxies & $5.48\pm0.83$ & 1.53 & 0.24 \\
26 & $10^{8}$ & $10^{8}$ & Random between 0 and 10 & $4.98\pm0.78$ & 1.50 & 0.23 \\
27 & $10^{8}$ & $10^{8}$ & 100 for all galaxies & $5.84\pm0.79$ & 1.55 & 0.22 \\
28 & $10^{8}$ & $10^{8}$ & Random between 0 and 100 & $5.68\pm0.82$ & 1.53 & 0.23 \\
29 & $10^{8}$ & $10^{8}$ & 1000 for all galaxies & $5.62\pm0.82$ & 1.59 & 0.27 \\
30 & $10^{8}$ & $10^{8}$ & Random between 0 and 1000 & $5.58\pm0.88$ & 1.57 & 0.26 \\
31 & $10^{9}$ & $10^{8}$ & 10 for all galaxies & $1.87\pm0.51$ & 1.37 & 0.26 \\
32 & $10^{9}$ & $10^{8}$ & Random between 0 and 10 & $2.33\pm0.58$ & 1.38 & 0.25 \\
33 & $10^{9}$ & $10^{8}$ & 100 for all galaxies & $1.11\pm0.40$ & 1.09 & 0.47 \\
34 & $10^{9}$ & $10^{8}$ & Random between 0 and 100 & $1.28\pm0.42$ & 1.19 & 0.37 \\
35 & $10^{9}$ & $10^{8}$ & 1000 for all galaxies & $1.94\pm0.50$ & 0.68 & 0.70 \\
36 & $10^{9}$ & $10^{8}$ & Random between 0 and 1000 & $1.31\pm0.40$ & 0.88 & 0.66 \\
            \noalign{\smallskip}
            \hline
         \end{tabular}
     $$ 
   \end{table}

\clearpage

   \begin{table}
      \caption[]{Averaged values of $f$(H$\alpha$)/$f$(UV) and $\log
b_{\mbox{\scriptsize obs}} /b_{\mbox{\scriptsize model}}$ for the various analyzed subsamples. Numbers in
parenthesis correspond to one standard deviation. 
}
         \label{tabmedia}
     $$ 
         \begin{tabular}{lcccc}
            \hline
            \noalign{\smallskip}
Subsample & Num. gal. & $\left< def(\mbox{H{\sc i}}) \right>$ & $\left<\log
f(\mbox{H}\alpha)/f(\mbox{UV})\right>$ & $\left<\log
b_{\mbox{\scriptsize obs}}/b_{\mbox{\scriptsize model}}\right>$ \\
            \noalign{\smallskip}
            \hline
            \noalign{\smallskip}
{\bf \em Reference} & {\bf 57} & {\bf $-$0.03(0.28)} & {\bf 1.11(0.24)} & {\bf 0.00(0.41)} \\
{\em Interacting} & 3 & 0.20(0.05) & 1.25(0.07) & 0.55(0.41) \\
{\em Asymmetric} & 6 & 0.11(0.27) & 1.24(0.20) & $0.01$(0.34) \\
{\em Deficient} & 27 & 0.66(0.18) & 1.25(0.27) & $0.07$(0.50) \\
            \noalign{\smallskip}
            \hline
         \end{tabular}
     $$ 
   \end{table}

\clearpage

   \begin{table}
      \caption[]{Average star formation activity in subsamples satisfying various selection criteria. 
}
         \label{bias}
     $$ 
         \begin{tabular}{lccc}
            \hline
            \noalign{\smallskip}
Observational & \multicolumn{3}{c}{$\left< EW(\mbox{H}\alpha + \mbox{[N{\sc ii}]}) \right>$} \\
constraint & \multicolumn{3}{c}{(\AA)} \\
            \noalign{\smallskip}
            \hline
            \noalign{\smallskip}
 & All & $def(\mbox{H{\sc i}}) < 0.4$ & $def(\mbox{H{\sc i}}) \geq 0.4$ \\
            \noalign{\smallskip}
            \hline
            \noalign{\smallskip}
None & 17 & 22 & 10 \\
FIR det. & 19 & 24 & 11 \\
UV det. & 21 & 28 & 12 \\
FIR \& UV det. & 23 & 31 & 13 \\
H$\beta$ det. & 27 & 29 & 20 \\
            \noalign{\smallskip}
            \hline
         \end{tabular}
     $$ 
   \end{table}

\clearpage


\clearpage

\clearpage

   \begin{figure}[t]
   \centering
 \includegraphics[width=8.5cm]{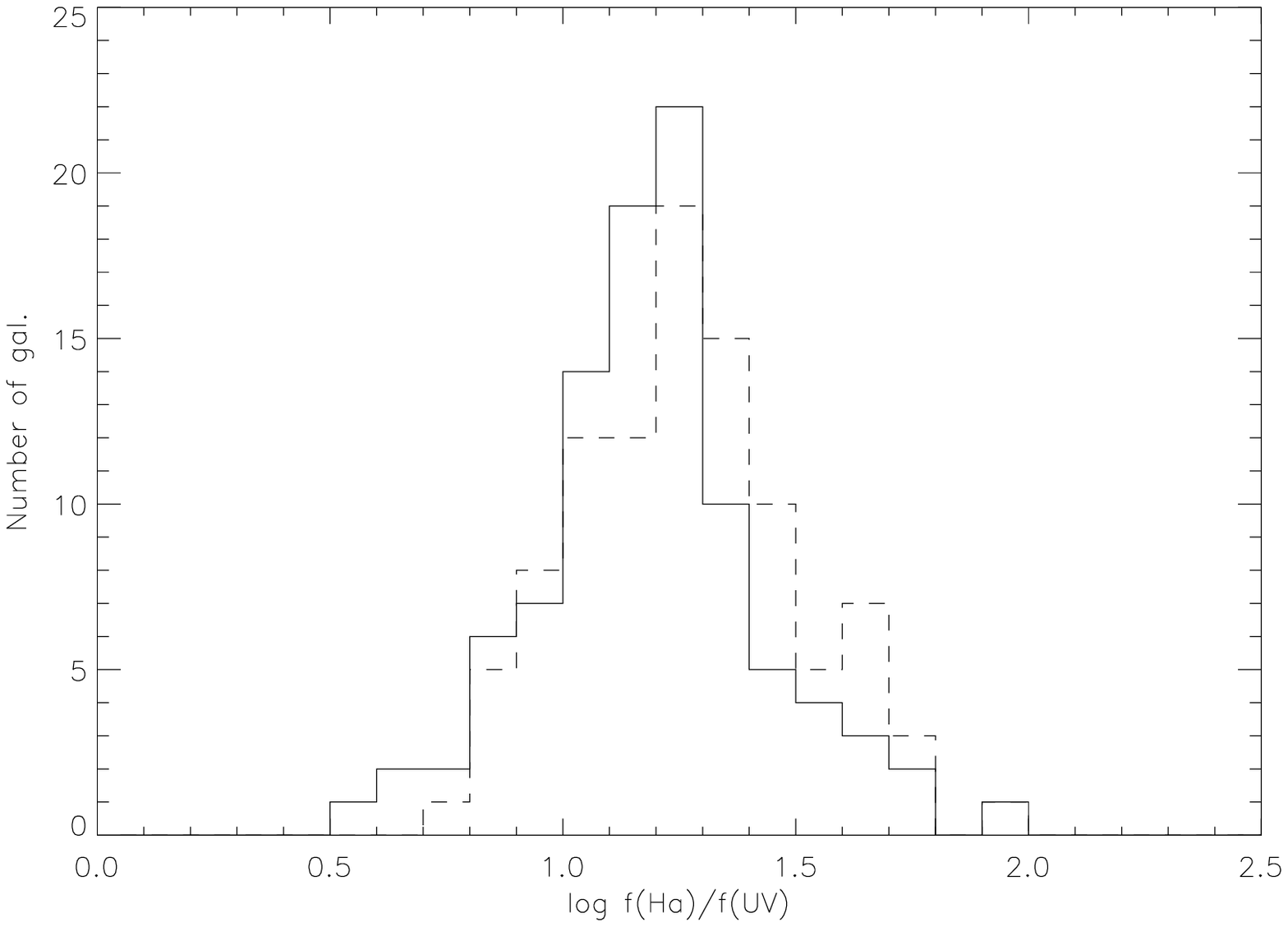}
 \includegraphics[width=8.5cm]{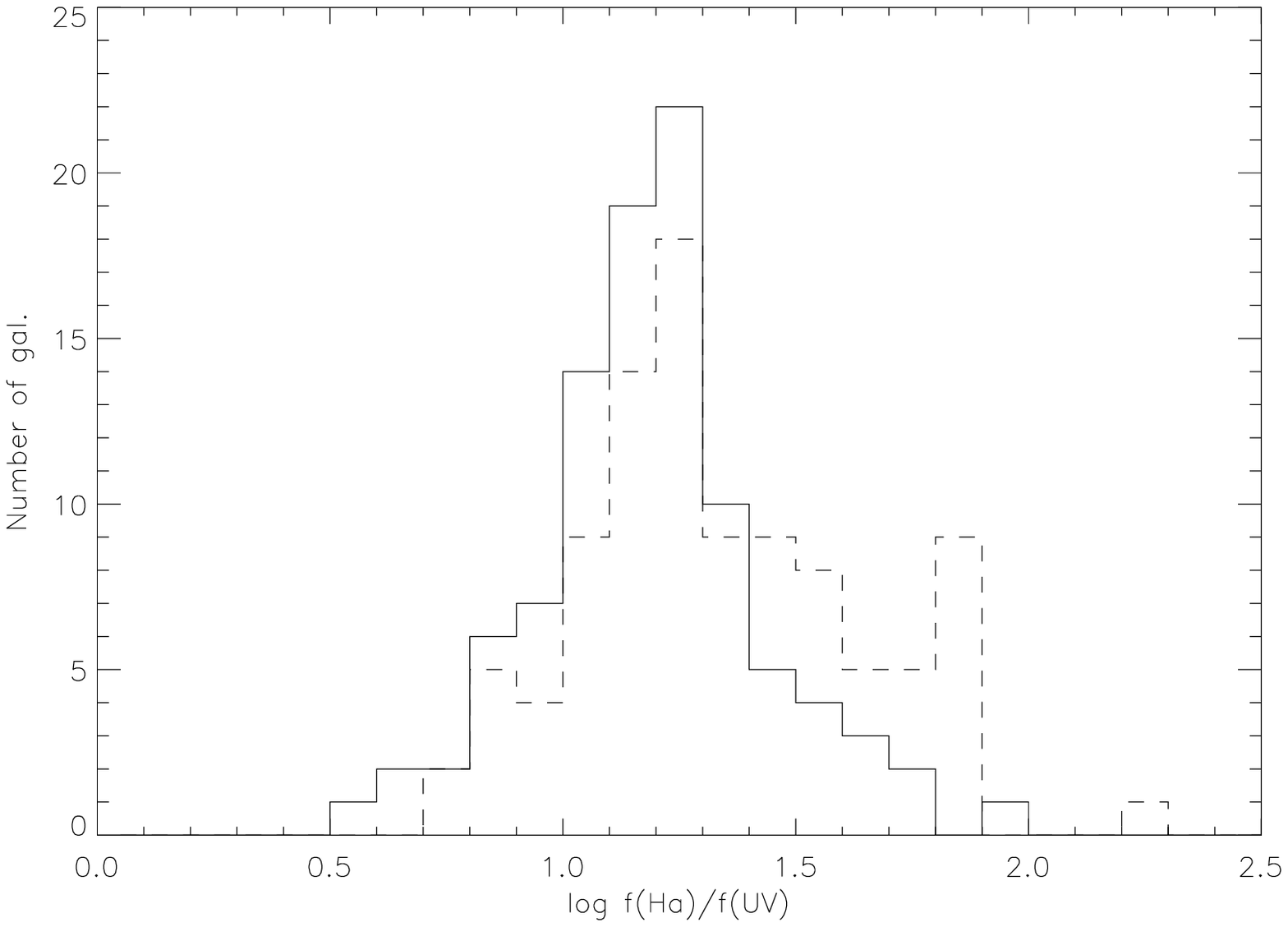}
 \includegraphics[width=8.5cm]{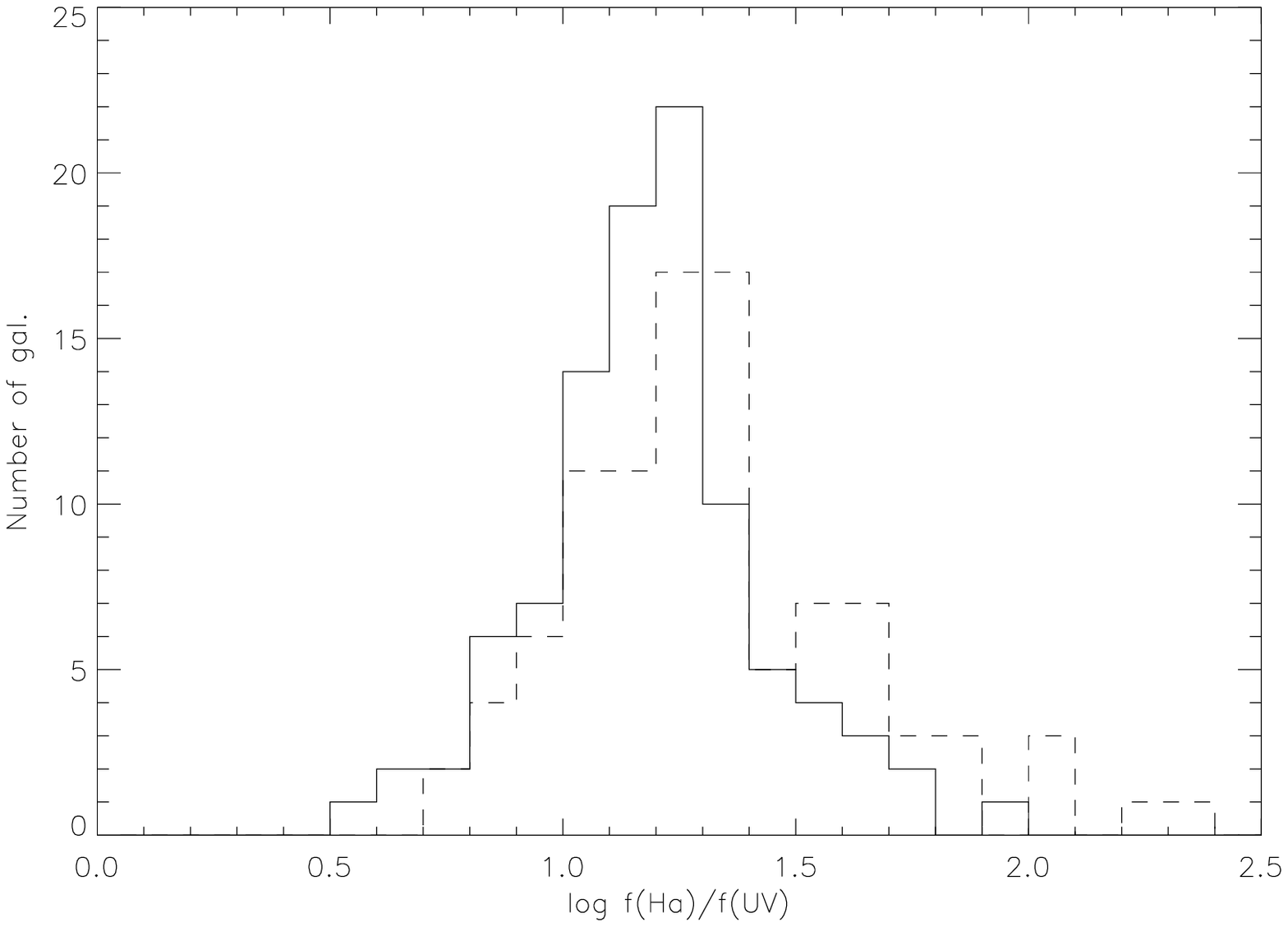}
 \includegraphics[width=8.5cm]{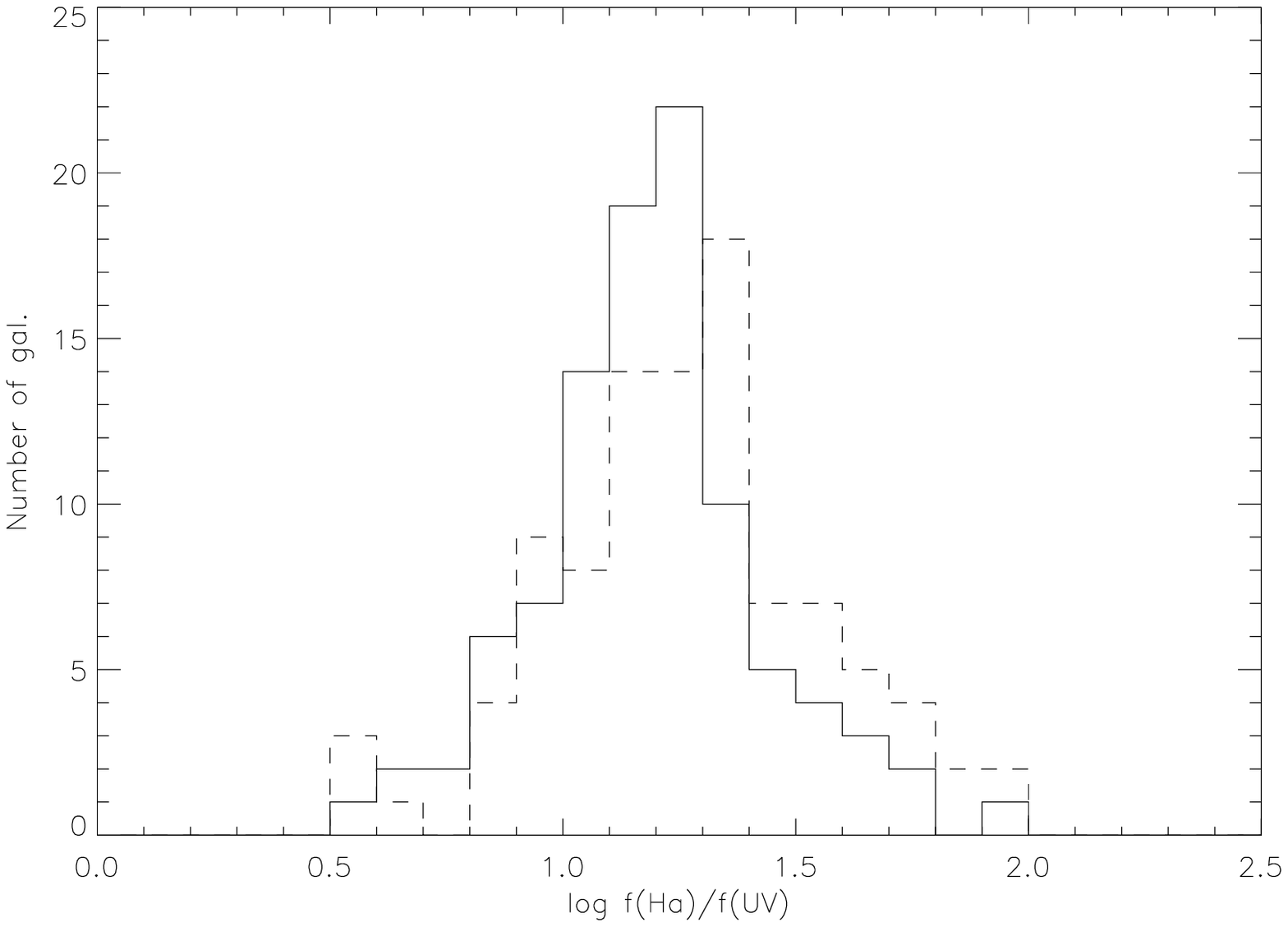}
 \includegraphics[width=8.5cm]{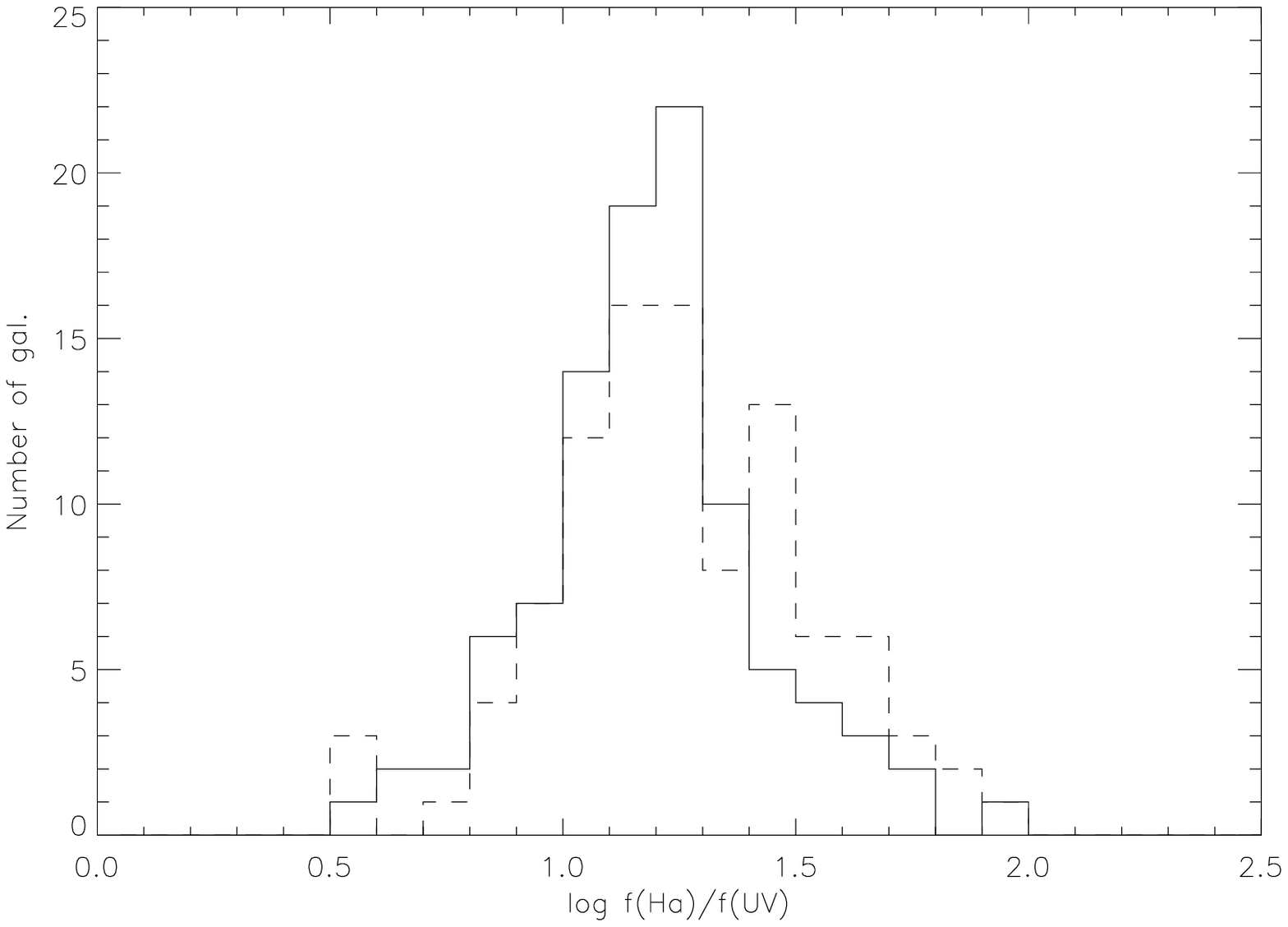}
 \includegraphics[width=8.5cm]{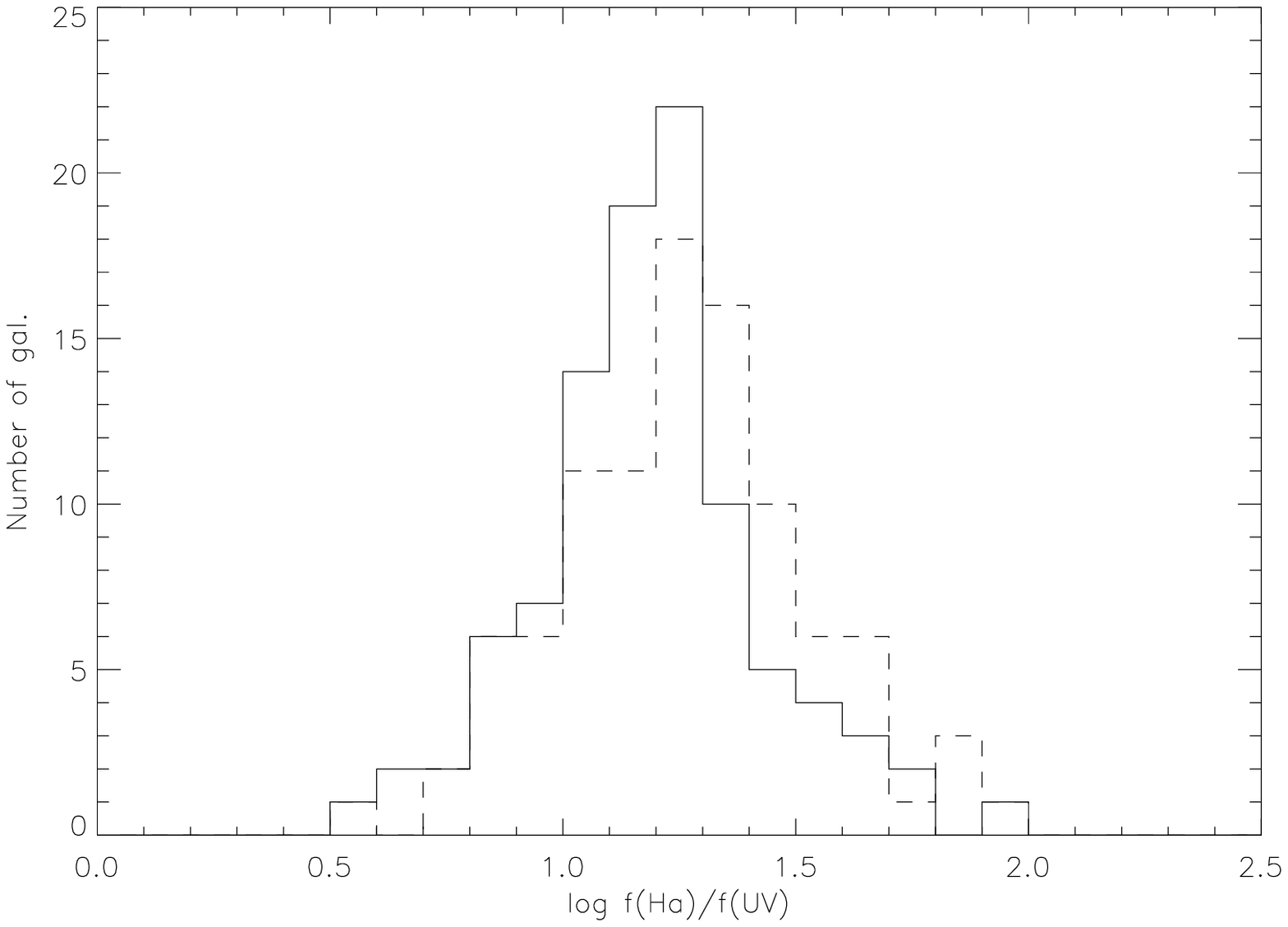}
      \caption{
Histograms of $f$(H$\alpha$)/$f$(UV) for six different simulations of
scenarios~9 and 12 (dashed lines). 
The observed histogram is given with solid lines.}
         \label{dt10}
   \end{figure}


   \begin{figure}[t]
   \centering
 \includegraphics[width=17cm]{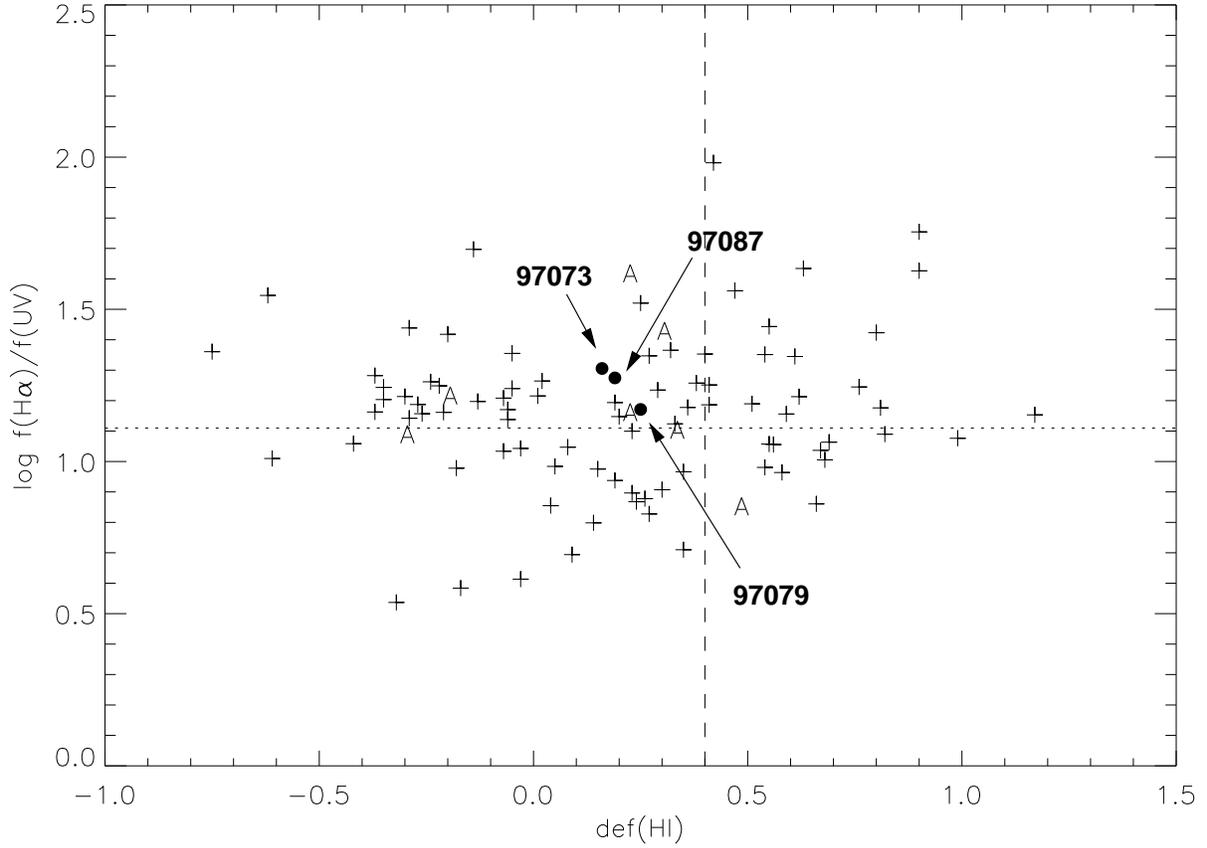}
      \caption{
The relationship between $\log$~$f$(H$\alpha$)/$f$(UV) and the H{\sc i} deficiency.
Interacting galaxies are marked with filled dots. 
Galaxies with asymmetric H{\sc i} profiles are labeled with
``A''. The
short-dashed horizontal line corresponds to the average value of
$f$(H$\alpha$)/$f$(UV) for the reference sample. The dashed vertical line
corresponds to $def(\mbox{H{\sc i}}) = 0.4$: plusses with $def$(H{\sc i}) $\leq 0.4$ represent the reference sample.}
         \label{ha_uv_hidef}
   \end{figure}

\clearpage

   \begin{figure}[t]
   \centering
 \includegraphics[width=17cm]{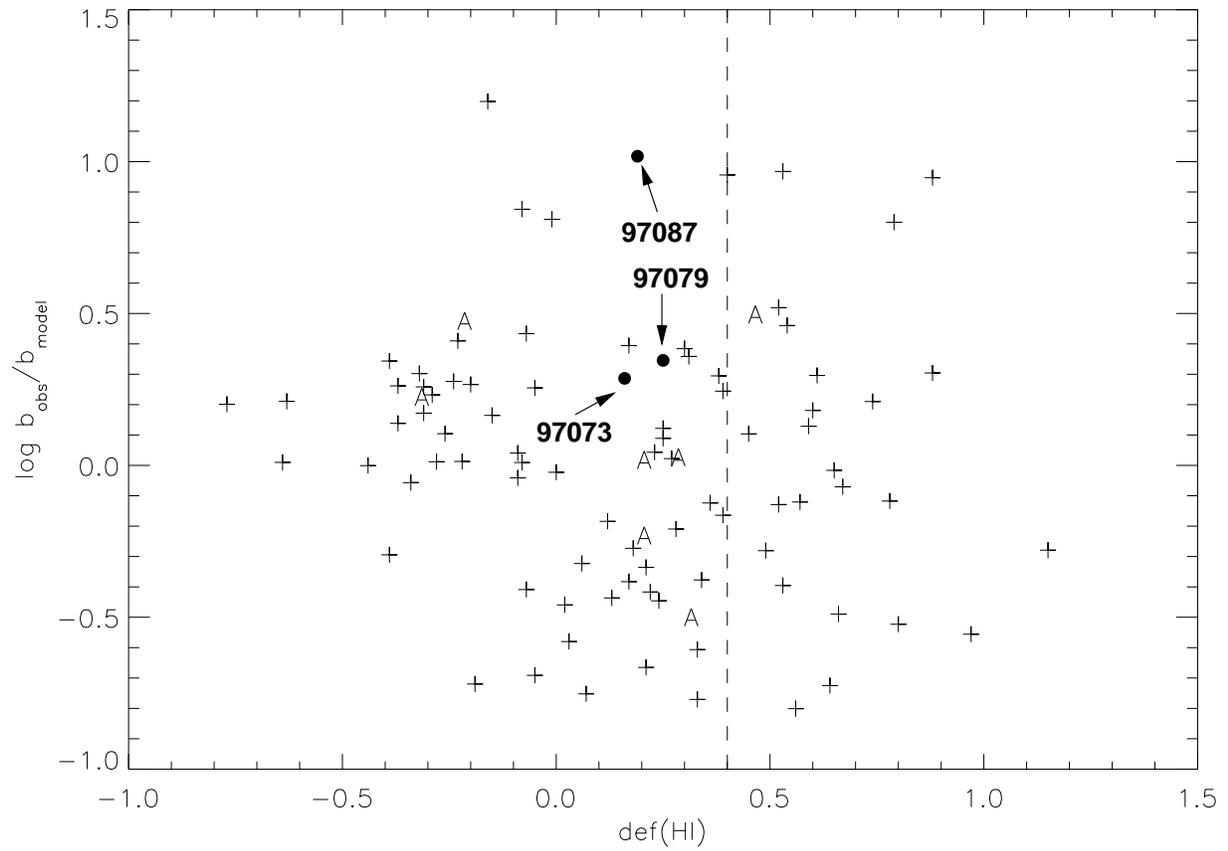}
      \caption{
The relationship between $\log b_{\mbox{\scriptsize obs}}/b_{\mbox{\scriptsize model}}$ and the H{\sc i} deficiency.
The dashed vertical line
corresponds to corresponds to $def(\mbox{H{\sc i}}) = 0.4$. Symbols as in fig.~\ref{ha_uv_hidef}.}
         \label{dbirth_hidef}
   \end{figure}

\newpage

\clearpage

\onecolumn

\appendix

\section{The intensity of the star formation bursts}

The $b$ parameter is the ratio of the recent to the total SFRs 
over the whole life of a galaxy as defined by Kennicutt et al. (1994).
If the SFR of a galaxy as a function of time is known, then: 
\begin{equation}
\label{b}
b = \frac{\mbox{SFR}(t,\tau) \times t}{\int_{0}^{t} \mbox{SFR}(t',\tau)dt'}
\end{equation}
where $t$ is the current epoch and the galaxies are assumed to be
formed at $t' = 0$. 
If a simple exponential SF history is assumed:
\begin{equation}
\mbox{SFR}(t,\tau) = \mbox{SFR}_{0} e^{-t/\tau}
\label{exp}
\end{equation}
the $b$ parameter can be expressed, following Boselli et al. (2001) as:
\begin{equation}
b_{\mbox{\scriptsize model}} = \frac{t \times e^{-t/\tau}}{\tau (1 - e^{-t/\tau})}
\label{bmod}
\end{equation}
These authors also report an empirical relationship:
\begin{equation}
\label{tau_h}
\log L_{\mbox{\scriptsize H}} = -2.5 \times \log \tau +12
\end{equation}
that, together with eq.~\ref{bmod}, provides the link between the $b$ parameter 
and the $H$-band luminosity of a galaxy, 
in the case of an exponential SFH of eq.~\ref{exp}.


An independent way to obtain the value of $b$ from purely
observational considerations is as following Boselli et al. (2001):
\begin{equation}
b_{\mbox{\scriptsize obs}} = \left( \frac{L_{\mbox{\scriptsize H}\alpha}}{10^{41}}\right) 
\times 0.26 \times \left( \frac{t}{L_{\mbox{H\scriptsize }}} \right)
\end{equation}
where $t$ is as in eq.~\ref{b} in yr, and $L_{\mbox{\scriptsize
H}\alpha}$ and $L_{\mbox{\scriptsize H}}$ are the H$\alpha$ and $H$-band luminosities
respectively.

The comparison of $b$ obtained from the average
empirical relationship between $\tau$ and $L_{\mbox{\scriptsize H}}$ (i.e. $b_{\mbox{\scriptsize model}}$),
and from $L_{\mbox{\scriptsize H}}$ and $L_{\mbox{\scriptsize H}\alpha}$
(i.e. $b_{\mbox{\scriptsize obs}}$) should reflect the
deviations from a smooth evolution on timescales  of the
order of $3 \times 10^{6}$~yr. 

\clearpage

\section{The error budget}

This appendix is aimed at estimating the total error
budget of the $f$(H$\alpha$)/$f$(UV) ratio as computed from our data.
We adopt the following expression for the $f$(H$\alpha$)/$f$(UV) ratio:
\begin{displaymath}
\log f(\mbox{H}\alpha)/f(\mbox{UV}) = \log f_{0}(\mbox{H}\alpha) \pm
\Delta f_{0}(\mbox{H}\alpha) - 
\log \left[ 1 + \frac{I(\mbox{H}\alpha)}{I(\mbox{[N{\sc
ii}]})}  \times \frac{1 \pm\Delta I(\mbox{H}\alpha)}{1 \pm\Delta I(\mbox{[N{\sc
ii}]})} \right] - \left(\frac{1}{0.335} - 1 \right) \times
\end{displaymath}
\begin{displaymath}
\times \log \left[
\frac{I(\mbox{H}\alpha)}{I(\mbox{H}\beta)}
\frac{1}{2.87} \times
\frac{1 \pm \Delta I(\mbox{H}\alpha)}{1 \pm
\Delta I(\mbox{H}\beta_{\mbox{\scriptsize emi}} \pm \Delta I(\mbox{H}\beta_{\mbox{\scriptsize abs}})} \right] -
\log f_{0}(\mbox{UV}) \pm \Delta f_{0}(\mbox{UV}) - 
\end{displaymath}
\begin{equation}
-0.466 -
\log \left[ \frac{f_{0}(\mbox{Far-IR})}{f_{0}(\mbox{UV})} \times \frac{1 \pm \Delta
f_{0}(\mbox{Far-IR})}{1 \pm \Delta f_{0}(\mbox{UV})} \right]  - 
0.433 \times \log \left[
\frac{f_{0}(\mbox{Far-IR})}{f_{0}(\mbox{UV})} \times \frac{1 \pm \Delta
f_{0}(\mbox{Far-IR})}{1 \pm \Delta f_{0}(\mbox{UV})} \right]^{2} 
\label{ecuerror}
\end{equation}
where,
\begin{itemize}
\item $f_{0}$(H$\alpha$), $f$(Far-IR) and $f_{0}$(UV) are the measured integrated
luminosities from imaging data in the corresponding passbands,
\item $\Delta f_{0}(\mbox{H}\alpha)$, $\Delta f_{0}(\mbox{Far-IR})$ and $\Delta
f_{0}(\mbox{UV})$ are the uncertainties of the H$\alpha$, Far-IR and UV fluxes,
\item $I$(H$\alpha$), $I$(H$\beta$) and $I$([N{\sc ii}]) are the fluxes of
the corresponding emission lines as measured from the optical spectra,
\item $\Delta I$(H$\alpha$), $\Delta I$(H$\beta_{\mbox{\scriptsize emi}}$) and $\Delta I$([N{\sc
ii}]) are the uncertainties on the fluxes of the
corresponding emission lines,
\item $\Delta I$(H$\beta_{\mbox{\scriptsize abs}}$) is the uncertainty on the flux of
the H$\beta$ absorption line.
\end{itemize}

The formula used to derive the extinction at 2000~\AA\ was taken from
Buat et al. (1999). In order to estimate our total error budget, we
run Monte-Carlo simulations of the distribution of 56 values with the
error budget shown in eq.~\ref{ecuerror}. The individual sources of
uncertainty were assumed to follow a gaussian distribution. 
The error sources are
listed in Table~\ref{formuerr}. For our simulations we assumed typical
values of $f_{0}(\mbox{Far-IR})/f_{0}(\mbox{UV}) = 1$,
$I(\mbox{H}\alpha)/I(\mbox{H}\beta) = 3$ and $I(\mbox{[N{\sc
ii}]})/I(\mbox{H}\alpha) = 0.2$. The simulated distributions turned
out to be fairly symmetric with typical dispersions of $\sigma = 0.27
\pm 0.03$~dex. The centers of the distributions showed typical
variations of $\pm 0.03$~dex.

   \begin{table}[!hb]
      \caption[]{The error sources
entering the computation of the uncertainty in the $f$(H$\alpha$)/$f$(UV) ratio.}
         \label{formuerr}
     $$ 
         \begin{tabular}{lc}
            \hline
            \noalign{\smallskip}
Uncertainty source & Estimated \\
            \noalign{\smallskip}
            \hline
            \noalign{\smallskip}
$\Delta f_{0}(\mbox{H}\alpha)$ & 15\% \\
$\Delta f_{0}(\mbox{Far-IR})$ & 15\% \\
$\Delta f_{0}(\mbox{UV})$ & 20\% \\
$\Delta I$(H$\alpha$) & 10\% \\
$\Delta I$(H$\beta_{\mbox{\scriptsize emi}}$) & 10\% \\
$\Delta I$(N[{\sc ii}]) & 15\% \\
$\Delta I$(H$\beta_{\mbox{\scriptsize abs}}$) & 20\% \\
            \noalign{\smallskip}
            \hline
         \end{tabular}
     $$ 
   \end{table}

\end{document}